\documentclass[12pt,titlepage]{article}

\usepackage{epsfig}
\usepackage{amsmath}
\usepackage{lscape}
\usepackage{amsmath,epsf,color,colordvi,shadow,pifont}
\usepackage[hypertex]{hyperref}

\renewcommand{\theequation}{\arabic{section}.\arabic{equation}}

\begin{document}

\begin{titlepage}
\begin{center}
\hfill{CERN-PH-TH-2014-111}\\
\hfill{CCTP-2014-10~~~~~~~~~~~~~}\\
\hfill{CCQCN-2014-29~~~~~~~~~}

\vskip 2cm

{\huge {\bf  Universality classes for models of inflation}}
 \\
 ~\\

\vskip 1cm

{\bf\large P. Bin\'etruy$^{ab}$\footnote{binetruy@apc.univ-paris7.fr}, E.
Kiritsis$^{acd}$, J. Mabillard$^{ab}$, M. Pieroni$^{ab}$, C. Rosset$^a$} \\
~\\
~\\
{\em ${}^a$ AstroParticule et Cosmologie,
Universit\'e Paris Diderot, CNRS, CEA, Observatoire de Paris, Sorbonne Paris
Cit\'e, F-75205 Paris Cedex 13} \\
{\em ${}^b$ Paris Centre for Cosmological Physics, F-75205 Paris Cedex 13}\\
{\em ${}^c$ \href{http://hep.physics.uoc.gr/}
 {Crete Center for Theoretical Physics}, Department of Physics, University of Crete
 71003 Heraklion, Greece}\\
{\em ${}^d$ \href{http://wwwth.cern.ch/}{Theory Group, Physics Department, CERN}, CH-1211, Geneva 23, Switzerland }

\end{center}

\vskip 1cm
\centerline{ {\bf Abstract}}

\indent
We show that the cosmological evolution of a scalar field in a potential
can be obtained from a renormalisation group equation. The slow roll regime of
inflation models is understood in this context as the slow evolution close to
a fixed point, described by the methods of renormalisation group.
This explains in part the universality observed in the predictions of a
certain number of inflation models. We illustrate this behavior on a certain
number of examples and discuss it in the context of the AdS/CFT correspondence.
\vfill
\end{titlepage}
\def\noi{\noindent}
\def\sq{\hbox {\rlap{$\sqcap$}$\sqcup$}}
\def\1{{\rm 1\mskip-4.5mu l} }
\def\gsim{\mathrel{\rlap{\lower4pt\hbox{\hskip1pt$\sim$}} \raise1pt\hbox{$>$}}}
\def\lsim{\mathrel{\rlap{\lower4pt\hbox{\hskip1pt$\sim$}} \raise1pt\hbox{$<$}}}
\def\be{\begin{equation}}
\def\ee{\end{equation}}
\def\bea{\begin{eqnarray}}
\def\eea{\end{eqnarray}}
\def\sp{\;\;\;,\;\;\;}
\def\e{\epsilon}
\def\l{\lambda}
\tableofcontents


\section{Introduction}

One of the cornerstones of early universe cosmology is inflation.
Over the years and especially recently, there has been an enormous amount of
data shaping our understanding of that period. Although the main lines of the
physics begin to become clearer, we seem to be still far from pinning down a
concrete (and hopefully theoretically well-motivated) model of inflation.
So far, theoretical prejudice dominates; especially there are good reasons to
believe that we do not understand very well the gravitational interaction and
its relation with the rest of the fundamental interactions.

Until now, inflationary model building relied on identifying inflationary
potentials, motivated by various principles of potential fundamental theories
(supergravity, string theory, etc.), or phenomenological considerations.
The recent results of the Planck mission \cite{Ade:2013uln} have successfully
constrained or ruled out a large set of models. More recently, the BICEP
results \cite{BICEP} have shown the outstanding scientific potential of B-mode
polarization.

It is clear however that such a landscape of models is not a good set of
local maps in order to identify what is possible in inflation, and
that many of the
descriptions are redundant. The reason is that the main properties of inflation
(near scale invariance and power spectra) depend on only few of the
details of inflationary potentials. A recent and topical  example of such a
degeneracy is  the connection between the original $R^2$ model
\cite{Starobinsky:1980te} and Higgs
inflation \cite{Bezrukov:2007ep}.

In order to go further, it is
important to classify inflationary  models in a systematic way in order to
understand the discriminative potential of cosmological observations.
Efforts in this direction have recently been made, in order to catalogue
inflationary models using other data than the scalar potential.  S. Mukhanov
\cite{Mukhanov:2013tua} has
proposed to consider a specific class of models representative of the most
successful inflationary models compatible with existing cosmological data: he
chooses an equation of state of the type
\be
{(p+\rho)\over \rho} ~~\sim ~~
{\beta\over (N+1)^\alpha}
\ee
 where $p$ and $\rho$ are the pressure and energy density
associated with the inflation field, and $N$ the number of $e$-foldings
to parametrize the evolution from a (almost) constant vacuum energy ($p=-\rho$
at early times i.e. $N$ large) to a radiation-dominated universe ($p$ of the
order of $\rho$ at late times i.e. small $N$); $\alpha$ and $\beta$ are
parameters discriminating the models. In \cite{Roest} and \cite{GB} this
proposal has been pushed further by enlarging the possible cases considered.

These  proposals however are still ``frame dependent". Our purpose is to
provide universality classes of models of inflation by using the most basic
property of inflation: the scaling invariance that, although approximate, is
always present. This scaling invariance and ideas originating in AdS/CFT tie
together the principle we are looking for with ideas of universality in
Quantum Field Theory (QFT) \cite{Kiritsis}.

A connection between the original AdS/CFT correspondence and cosmology, named  ``mirage cosmology", was suggested in \cite{mirage}, via a universe-brane motion in the near horizon region of D$_3$ branes.
It has been independently  suggested in \cite{strominger} that there could be a correspondence between physics in de Sitter space and a dual (pseudo) Conformal Field Theory. In \cite{McFadden} it was argued that  cosmological evolution
equations have a formal resemblance with similar equations in Anti de Sitter
space, describing holographic Renormalization Group (RG) flows in QFTs with a
holographic dual. This correspondence was extended further in \cite{Kiritsis}
where new classes of holographic CFTs have been included and interpreted:
namely asymptotically free CFTs. These  correspond in the cosmological context
to Asymptotically Flat Inflationary Models (AFIM), that include as special
cases, the Starobinsky model and Higgs inflation.

The correspondence depicted in \cite{McFadden} and \cite{Kiritsis} is not
necessarily an equivalence, but at this stage can be taken as a rather
remarkable similarity between the QFT and the inflationary setup. As such it
suggests that the proper framework for defining universality classes for
inflationary models is the Wilsonian picture of fixed points (that here
corresponds to exact deSitter solutions), scaling regions (that in cosmology
corresponds to inflationary periods), and critical exponents (that in cosmology
corresponds to the scaling exponents of power spectra).
It is well known from QFT that these are the optimal universal
characterisations of nearly scale invariant dynamics.

In what follows, we use the Hamilton-Jacobi formalism of Salopek and Bond
\cite{Salopek:1990jq}
to parametrize the solutions of the cosmological evolution of a scalar field
in its potential in terms of a superpotential. The evolution can be viewed
as a standard renormalisation group equation. This
allows to discuss classical inflation scenarios as the slow evolution of
a system approaching or leaving  a critical (fixed) point.
The parametrization of the associated $\beta$ functions provides a way of
defining the universality classes. Unlike the standard case of perturbative
QFT, the $\beta$ functions that we will consider here may be of a
non-perturbative nature.

Although in this paper we discuss single field models of inflation, our
classification can be extended also to multifield models, as explained in
\cite{Kiritsis},\cite{Bourdier}. More phenomena are possible in this more
general case, but we will not discuss them in this paper.

The structure of this paper is as follows. In section \ref{sect:1}, we present
the Hamilton-Jacobi formalism that we will be using and discuss in this context
the analysis of Mukhanov \cite{Mukhanov:2013tua} mentionned above. Section
\ref{sect:2} defines our different classes of models and gives their main
predictions. In Section \ref{sect:4}, we make the connection with quantum
field theories using the AdS/CFT correspondence, and rediscuss from this
perspective the different classes introduced. In Section \ref{sect:5},
we discuss inflation models which interpolate between two classes, such as
natural inflation or hilltop inflation. Finally, Section \ref{sect:6} gives
our conclusions.

\section{Cosmological evolution of a scalar field in a potential\label{sect:1}}
\setcounter{equation}{0}

We recall the equations governing the cosmological evolution of a homogeneous
classical field $\phi(t)$ in its potential $V(\phi)$:
\begin{equation}
\label{action}
{\cal S}= \int d^4x \sqrt{-g} \left[ -{m_{_P}^2 \over 2} R - { 1 \over 2}
\partial^\mu \phi \partial_\mu \phi - V(\phi) \right] \ .
\end{equation}

 We consider a Friedmann
Lema\^itre Robertson Walker universe, with scale factor $a(t)$, and we assume
flatness for simplicity i.e. $ds^2 = -dt^2 + a(t)^2 (dr^2 + r^2 d\Omega^2)$. One
obtains from Einstein's equations:
\begin{eqnarray}
H^2 \equiv {{\dot a}^2   \over a^2 } &=& {\kappa^2 \over 3} \rho_\phi \ ,
\label{Hubble}\\
 {\dot a}^2 + 2 a {\ddot a}  &=& - \kappa^2 a^2 p_\phi \ ,
\label{acceleration}
\end{eqnarray}
where $\kappa^2 = 8 \pi G_{_N} = 8\pi/M_{_P}^2$ and
\begin{eqnarray}
p_\phi &=& {1 \over 2} \dot \phi^2 - V( \phi) \quad , \label{pphi} \\
\rho_\phi &=& {1 \over 2} \dot \phi^2 + V( \phi)  \label{rhophi}
\end{eqnarray}
are respectively the pressure and energy density associated with the field
$\phi$. The equation of motion for this field, namely
\begin{equation}
\label{phieom}
\ddot \phi + 3 H \dot \phi = - {dV \over d \phi} \ ,
\end{equation}
is consistent with the conservation of $\rho_\phi$ in the expanding universe:
\begin{equation}
\label{rhoconserv}
\dot \rho_\phi = -3H (p_\phi + \rho_\phi) \ .
\end{equation}

The Hamilton-Jacobi approach of Salopek and Bond \cite{Salopek:1990jq} (see
also \cite{Lidsey:1995np}) provides a very useful integral of motion of
the system of equations that we just wrote. Indeed, assuming that the evolution
of the scalar field is (piece-wise) monotonic, we may consider that the $\phi$
field provides a clock by just inverting $\phi(t)$ into $t(\phi)$. Then the
Hubble parameter may be considered as a function of $\phi$ and we write
\begin{equation}
\label{superpotential}
H = {\dot a \over a} \equiv -{1 \over 2} W(\phi) \ .
\end{equation}
Differentation gives $(\ddot a/a)-(\dot a/a)^2 = -(1/2) W_{,\phi} \dot \phi$
whereas (\ref{Hubble}), (\ref{acceleration}) yield
$(\ddot a/a)-(\dot a/a)^2 = -
\kappa^2 (p_\phi + \rho_\phi)/2 = - \kappa^2 \dot \phi^2/2$. We thus have
\begin{equation}
\label{dotphi}
\dot \phi ={1 \over \kappa^2} W_{,\phi} \ .
\end{equation}
Then, using (\ref{rhophi}) and (\ref{Hubble}), one may write the potential
as
\begin{equation}
\label{potsuperpot}
2\kappa^2 V = {3 \over 2} W^2 - {1 \over \kappa^2} W_{,\phi}^2 \ ,
 \end{equation}
which form leads to call {\em superpotential} the function $W(\phi)$
(by reference to a similar parametrisation of the potential in the context of
supersymmetric quantum mechanics; see for example \cite{Binetruy:2006ad}).
The $\phi$ field equation of motion (\ref{phieom}), or equivalently the
equation of conservation (\ref{rhoconserv}), is then automatically satisfied.

The original system of equations needs 3 constants of integration, two for the
second order equation for the scalar and one for the first order equation for
the scale factor. Once we introduce the superpotential, the first order
equations (\ref{superpotential}) and (\ref{dotphi}) have two constants of
integration. The third ones is hidden in the first order differential equation
(\ref{potsuperpot}) that determines the superpotential in terms of the
potential. Therefore, for a single potential, there is an infinite family of
superpotentials satisfying (\ref{potsuperpot}). Only a discrete number
(typically one) of the solutions to (\ref{potsuperpot}) leads to regular
solutions of (\ref{superpotential}) and (\ref{dotphi}), \cite{ihqcd,T}.
All others have curvature singularities.

We note for future use that pressure and energy density are expressed in terms
of the superpotential as follows:
\begin{eqnarray}
\rho_\phi &=& {3 \over 4 \kappa^2} W^2 \ , \label{rhoW} \\
p_\phi + \rho_\phi &=& {1 \over \kappa^4} W_{,\phi}^2 \ . \label{p+rhoW}
\end{eqnarray}
which indicates that the null energy condition is always satisfied.
Using (\ref{superpotential}) and (\ref{dotphi}), we have
\begin{equation}
\label{RG}
\kappa {d\phi \over d \ln a} = {\kappa \dot \phi \over H} = -{2 \over \kappa}{W_{,\phi}
\over W}(\phi) \ ,
\end{equation}
which has exactly the form of a renormalisation group equation giving the
evolution of a renormalized coupling $g$ in terms of the renormalization scale
$\mu$, \cite{ihqcd}:
\begin{equation}
\label{RGtrad}
{d g \over d\ln \mu} = \beta (g) \ .
\end{equation}
Using (\ref{rhoW}) and (\ref{p+rhoW}), we note that the $\beta$-function in our
case is
\begin{equation}
\label{beta}
\beta(\phi) = -{2 \over \kappa}{W_{,\phi} \over W}
= \pm \sqrt{3{p_\phi + \rho_\phi \over \rho_\phi}}\ .
\end{equation}
Fixed points correspond to zeros of $W_{,\phi}$, which are extrema of the
potential $V(\phi)$ as can be seen from (\ref{potsuperpot}). They correspond to exact de Sitter solutions. An inflationary
period corresponds to the {\em slow motion} of the field away ($V_{,\phi\phi}<0$)
or towards ($V_{,\phi\phi}>0$) the fixed point. This  corresponds to the
classical distinction between a convex and a concave potential.

We note from (\ref{beta}) that, fixing the ratio $(p_\phi + \rho_\phi) /
\rho_\phi$ as a function of $\ln a$, or equivalently as the number of
$e$-foldings
\begin{equation}
\label{N}
N = - \ln (a/a_f)
\end{equation}
as in \cite{Mukhanov:2013tua} ($a_f$ is the value of $a$ at
the end of inflation), we are in a position to solve (\ref{RG}). For example,
the choice made by Mukhanov\footnote{This choice is special and does not contain all possible universality classes discussed in this paper.}  \cite{Mukhanov:2013tua}
\begin{equation}
\label{Mukh1}
\beta(\phi) = \pm \sqrt{3\beta}{1\over (N+1)^{\alpha/2}}
\end{equation}
gives using (\ref{RG}), with $\alpha \not = 2$
\begin{equation}
\label{Mukh2}
\phi = \phi_0 \mp {2 \over \kappa}{\sqrt{3\beta} \over 2-\alpha} (N+1)^{{2-
\alpha \over 2}} \ .
\end{equation}
This allows to express $N+1$ in terms of $\phi$ in the explicit form
(\ref{Mukh1}) of the  $\beta$-function $\beta(\phi)$. The superpotential is then
obtained by solving (\ref{beta}) i.e. $W_{,\phi}/ W = - \kappa^2 \beta(\phi)/2$:
\begin{eqnarray}
W &=& W_0 \exp \left[ -{3 \beta\over 2(\alpha-1)} X^{{\alpha-1\over \alpha-2}}
\right] \ , \quad X \equiv {\kappa^2 \over 12}{(2-\alpha)^2 \over \beta}
(\phi-\phi_0)^2 \ ,\label{Mukh21W} \\
V &=& {3\over 4\kappa^2} W_0^2 \left[1-{\beta\over 2}
X^{{\alpha \over \alpha-2}}\right]
\exp \left[ -{3 \beta \over \alpha-1}X^{{\alpha-1 \over \alpha-2}} \right] \ ,
\label{Mukh21}
\end{eqnarray}
where we assumed $\alpha \not = 1,2$ and $W_0$ is a constant (of mass
dimension $1$). For $\alpha = 1$, $W = C (\phi -\phi_0)^{3\beta}$ ($C$ constant)
and
\begin{equation}
\label{Mukh15}
V = {C^2 \over 2 \kappa^2} (\phi-\phi_0)^{6\beta} \left[{3\over 2} - {9 \beta^2
\over \kappa^2}{1 \over (\phi-\phi_0)^2}\right] \ .
\end{equation}
Finally, for $\alpha = 2$,
\begin{eqnarray}
W &=& W_0 \exp \left[ -{3 \beta \over 2} Y \right] \ ,
\quad Y \equiv \exp \left[\pm{\kappa \over \sqrt{3 \beta}}(\phi-\phi_0)\right]
\ ,\label{Mukh20W} \\
V &=& {3\over 4\kappa^2} W_0^2 \left[1-{\beta\over 2} Y^2\right] e^{-3\beta Y}
 \ .
\label{Mukh20}
\end{eqnarray}
Note that, for $Y \ll 1$, one has
\be
V \sim (3W_0^2/4\kappa^2)
\left[ 1-{3\beta\over 2} \exp\left(-{\kappa \over \sqrt{3 \beta}}(\phi-\phi_0)
\right)\right]^2\;.
\ee
One recovers
the Higgs inflation model \cite{Bezrukov:2007ep}.

\section{The Universality Classes}
\label{sect:2}
\setcounter{equation}{0}

We now classify the different  inflation models according to the behaviour
of the  $\beta$-function (\ref{beta}) close to the fixed point. The number of
e-foldings is simply
\begin{equation}
\label{Nbeta}	
N =-\kappa\int_{\phi_{f}}^{\phi}\frac{\mathrm{d}\phi'}{\beta(\phi')} \ .
\end{equation}
Assuming that we are close to the fixed point, we have $\beta(\phi) \ll 1$
in which case we have the simple expressions (for complete expressions, see
Appendix A) for the tensor-to-scalar ratio $r$,
the scalar and tensor spectral indices $n_s$ and $n_t$ and their
running\footnote{We note that, since $k$ refers to horizon crossing $k=aH$
during inflation, $d\ln k/d\phi = \kappa (1- \beta^2/2)/\beta$.} in terms of
$\beta(\phi)$ and its derivatives:
\begin{eqnarray}
\label{r}
r &=& 8 \beta^2 \ , \\
\label{ns}
n_{s}-1 &\simeq& -\left[\beta^{2}+\frac{2\beta'}{\kappa}\right] \ , \\
\label{nt}
n_{t} &\simeq& -\beta^{2} \ , \\
\label{dns/dlnk}
\alpha_s \equiv \frac{\mathrm{d}n_{s}}{\mathrm{dln}k} & \simeq &
-\frac{2}{\kappa}\beta^2\beta'-\frac{2}{\kappa^2}\beta\beta'' \ , \\
\label{dnt/dlnk}
\frac{\mathrm{d}n_{t}}{\mathrm{dln}k} &\simeq& -\frac{2}{\kappa}\beta^2\beta'
\ .
\end{eqnarray}
This can be expressed in terms of the (Hubble) slow roll
parameters\footnote{Following \cite{Lidsey:1995np}, we define the slowroll
parameters in terms of the field dependence of the Hubble parameter, in a way
consistent with the Hamilton-Jacobi approach (see (\ref{superpotential}): the
$H(\phi)$ function can just as well be replaced by the superpotential in
(\ref{epsilonH})-(\ref{xiH})).}
\begin{eqnarray}
\label{epsilonH}
\epsilon_H &\equiv& \frac{2}{\kappa^2}\left(\frac{H'}{H}\right)^{2}
=\frac{1}{2}\beta^2 \ , \\
\label{etaH}
\eta_H &\equiv& \frac{2}{\kappa^2}\frac{H''}{H}=\frac{1}{2}\beta^2-
\frac{\beta'}{\kappa} \ , \\
\label{xiH}
\xi_H^2 &\equiv& \frac{4}{\kappa^4}\left(\frac{H'H'''}{H^2}\right)=\frac{1}{4}
\beta^4-\frac{3}{2\kappa}\beta^2\beta'+\frac{1}{\kappa^2}\beta\beta'' \ .
\end{eqnarray}
to read
\begin{eqnarray}
\label{rH}
r = -8n_t &=& 16\epsilon_H \ , \\
\label{nSH}
n_{s}-1 &\simeq & 2\eta_H-4\epsilon_H \ , \\
\label{dnSdlnkH}
\frac{\mathrm{d}n_{s}}{\mathrm{dln}k} &\simeq & -8\epsilon_H^2
+10 \epsilon_H\eta_H-2\xi_H^2 \ , \\
\label{dnTdlnkH}
\frac{\mathrm{d}n_{t}}{\mathrm{dln}k} &\simeq & -4\epsilon_H^2
+4 \epsilon_H\eta_H \ .
\end{eqnarray}
Table  \ref{table} summarizes the different classes that we now discuss.

\begin{table}[htdp]
\begin{center}\begin{tabular}{c|c|c}
\textbf{Class} & \textbf{Name} &$\beta(\phi)$\\
\hline\hline
{\bf Ia(q)} & Monomial & $ \beta_{q}(\kappa\phi)^{q}$, $q>1$ \\
\hline
{\bf Ia(1)} & Linear & $ \beta_{1}(\kappa\phi)$ \\
\hline
{\bf Ib(p)} & Inverse Monomial &  $-\beta_p /(\kappa\phi)^p$, $p>1$ \\
\hline
{\bf Ib(1)} & Chaotic &  $-\hat \beta_1 /(\kappa\phi)$ \\
\hline
{\bf Ib(p)} & Fractional &  $-\hat \beta_p /(\kappa\phi)^p$, $0<p<1$ \\
\hline
{\bf Ib(0)} & Power Law &  $-\hat \beta_0$ \\
\hline\hline
{\bf II($\gamma$)} & Exponential & $-\beta\exp[-\kappa\phi]$
\end{tabular} \caption{Summary of the classes discussed in this Section.
}\label{table}
\end{center}

\end{table}

\subsection{Small field inflation ({\bf Ia})} \label{Ia}

Let us first consider the simplest class of models ({\bf Ia}), {\em where the
fixed point
is at a finite value $\phi_0$} and the  $\beta$-function has the following
expansion close to the fixed point:
\begin{equation}
\label{betaIa}
\beta(\phi) = \beta_q \left[\kappa(\phi - \phi_0)\right]^q  \ , \ \
q \in R \ .
\end{equation}
\vskip .3cm
$\bullet$ {\bf Monomial class: Ia(q)}, $q> 1$
\vskip .3cm
We first consider the case $q>1$\footnote{The case $q<1$ does not correspond to
inflation since the second derivative of the potential is divergent at the
fixed point, as can be seen from  (\ref{VIa}).}. Then, from (\ref{beta}) and
(\ref{potsuperpot}), we obtain,
\begin{eqnarray}
W(\phi) &=& W_0 \exp \left\{- {\beta_q \over 2 (q+1)}  \left[\kappa(\phi -
\phi_0)\right]^{q+1}  \right\} \ , \label{WIa} \\
V(\phi) &\sim& {3 W_0^2 \over 4 \kappa^2} \left[1 -  {\beta_q \over q+1}
\left[\kappa(\phi - \phi_0)\right]^{q+1} + {\cal O}(\phi^{2q}) \right] \ ,
\label{VIa}
\end{eqnarray}
where $W_0 = -2H_i$, $H_i$ being the value of the Hubble parameter at the fixed
point.
We then compute the number of e-foldings:
\begin{equation}
\label{NIa}
N = {1 \over \beta_q (q-1) \left[\kappa(\phi - \phi_0)\right]^{q-1}} - \lambda
\ ,
\end{equation}
where we defined
\be
\label{equ:lambdadef}
\lambda  \equiv {1\over  \beta_q (q-1) \left[\kappa(\phi_f -
\phi_0)\right]^{q-1}}\;.
 \ee
 Assuming that typically $\beta(\phi_f) \sim 1$, we
obtain
\be
\lambda \sim {\beta_q^{-1/q}\over (q-1)}\;,
\ee
 maximally of order $1$. The  $\beta$-function can be expressed in terms of $N$:
\begin{equation}
\label{betaNIa}
\beta(N) = {1 \over \left[  \beta_q^{\frac{1}{q}} (q-1) (N+\lambda)
\right]^{\frac{q}{q-1}}}
\end{equation}
Neglecting $\lambda$, we have for the scalar spectral
index:
\be
\label{nSIa}
n_s -1 \sim - 2\beta'/\kappa = - 2 q \beta_q \left[\kappa(\phi -
\phi_0)\right]^{q-1}
\sim  - {2q\over q-1} {1 \over N} \ ,
\ee
\be
\label{aSIa}
\alpha_s\equiv {dn_s \over d \ln k} \sim - 2 \beta \beta'' / \kappa^2
= - 2 q (q-1) \beta_q^2 \left[\kappa(\phi - \phi_0)\right]^{2(q-1)}
=-{2q \over q-1} {1 \over N^2} \ ,
\ee
whereas the tensor to scalar ratio reads
\begin{equation}
\label{rIa}
r = 8 \beta^2 \sim {8 \over
\beta_q^{2/(q-1)} \left[(q-1) N\right]^{2q/(q-1)}} \ .
\end{equation}

\begin{figure}[h]
\begin{center}
\epsfig{file=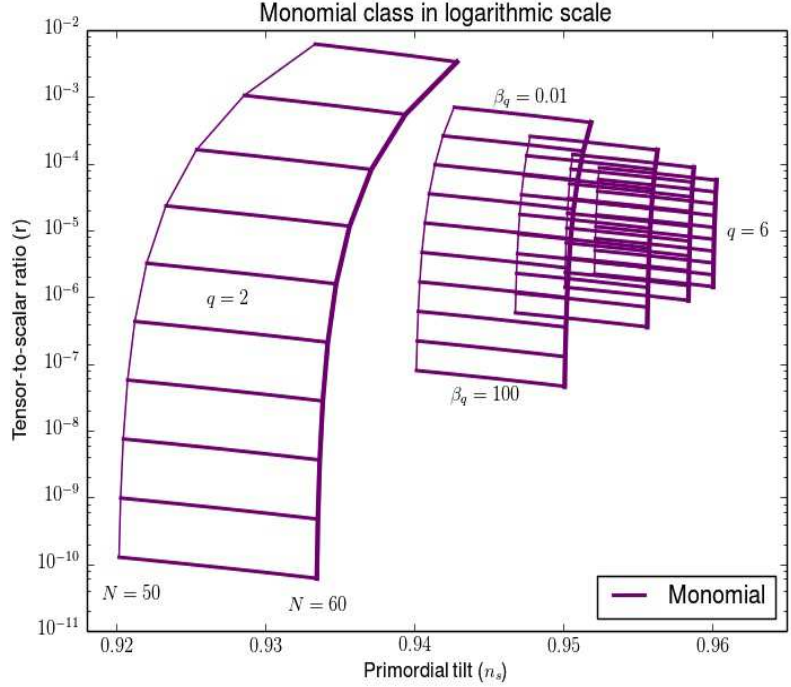,width=12cm,height=10cm}
\end{center}
\caption{Models of the monomial class ({\bf Ia(q)}, $q>1$) in the plot ($(r,
n_s$): values of $q$ are from left to right $=2,3,4,5,6$, and values of
$\beta_q$ (see (\ref{betaIa})) range from $10^{-2}$ (top) to $10^2$ (bottom)
in logarithmic spacing. Each horizontal segment corresponds to values of $N$
ranging from $50$ to $60$. The maximal relative error is $9 \cdot10^{-4}$.}
\label{figmonomial}
\end{figure}

We show in Figure \ref{figmonomial} the typical region covered by this class of
models in the plot $r$ vs $n_s$. In this Figure, as well as in similar ones
relative to other classes in the following subsections, we limit ourselves to
values of the parameters where we can safely consider that we remain
perturbatively close to the de Sitter regime, in agreement with the hypotheses
made in the previous Section. We give in each instance the maximal relative
error made when approximating the complete expressions to lowest order.\footnote{For more details on the procedure we have used see the end of Appendix \ref{App:A}}

We note here that the three experimentally accessible parameters, $n_s,
\alpha_s, r$, depend on three other parameters, namely $q,N$ and $\beta_q$.
This implies that in this class, generically speaking,  one can adjust
independently the three cosmological parameters (if this is compatible with
our assumption of perturbativeness). This is illustrated in Figure
\ref{figmonomial} where we see that such models cover a large area of the
parameter space. In particular there are many choices that reproduce the
preferred value $1-n_s\simeq 0.04$.

For completeness, we give here
the slow roll parameters in terms of $N$, to leading order:
\begin{equation}
\label{reltoNIa}
\eta_H = - {q\over q-1} {1 \over N} \ , \ \ \ \epsilon_H = {1 \over 2
\beta_q^{2/(q-1)} \left[(q-1) N\right]^{2q/(q-1)}} \ , \ \ \ \xi_H^2 =
{q \over q-1} {1 \over N^2} \ .
\end{equation}
We note that the generalized slowroll parameters \cite{Lidsey:1995np} have the
following behaviour:
\begin{equation}
\label{epsilonn}
\epsilon_n \equiv {2 \over \kappa^2} \left[{(H')^{n-1} H^{(n+1)} \over
H^n}\right]^{1/n}
\sim \left[\kappa(\phi-\phi_0)\right]^{q-1} \sim N^{-1} \ .
\end{equation}
In particular, they remain small close to the fixed point, as long as $q>1$.

We note from (\ref{betaNIa}) that this class of models corresponds to
$\alpha = {2q\over (q-1)}$ in the
notations of (\ref{Mukh1}), with $\alpha > 2$.

\vskip .3cm
$\bullet$ {\bf Linear class: Ia(1)},
\vskip .3cm
A special case is obtained for $q=1$:
\begin{equation}
\label{betaIasp}
\beta(\phi) = \beta_1 \kappa (\phi-\phi_0) \ .
\end{equation}
The corresponding superpotential and potential are:
\begin{eqnarray}
\label{WIasp}
W(\phi) &=& W_0 \exp \left[ - {\beta_1 \over 4} \left[\kappa(\phi - \phi_0)
\right]^2 \right] \ ,  \\
\label{VIasp}
V(\phi) &=& {3 W_0^2\over 4 \kappa^2} \left[1 -  {1 \over 2} \beta_1 (1 +
{\beta_1 \over 3})   \left[\kappa(\phi - \phi_0)\right]^2 + {\cal O}\left(
\left[\kappa(\phi - \phi_0)\right]^4 \right) \right] \ .
\end{eqnarray}
The number of e-foldings is, following (\ref{Nbeta}),
\begin{equation}
\label{NbetaIasp}
N = -{1 \over \beta_1} \ln \left[\beta_1 \kappa (\phi - \phi_0)\right] \ ,
\end{equation}
where we have supposed that $\beta(\phi_f) \sim 1$. It follows that the
$\beta$-function is expressed in terms of $N$ as
\begin{equation}
\label{betaNIasp}
\beta(N) = e^{-N\beta_1} \ .
\end{equation}
We obtain for the tensor to scalar ratio and the scalar spectral index:
\begin{eqnarray}
\label{rIasp}
r &=& 8 \beta_1^2  \left[\kappa(\phi - \phi_0)\right]^2 = 8 e^{-2N\beta_1} \ , \\
\label{nSIasp}
n_s -1 &\sim&  - 2\beta_1  \ , \ \ \ \alpha_s = {dn_s \over d \ln k}
= - 2 \beta_1 e^{-2N\beta_1} \ .
\end{eqnarray}
The slow roll parameters read
\begin{eqnarray}
\epsilon_H &=&  {1 \over 2}  e^{-2N\beta_1} \ ,
\label{epsilonIasp} \\
\eta_H &=& - \beta_1   \ , \label{etaIasp}
\\
\xi_H^2 &=& - {3 \over 2} \beta_1  e^{-2N\beta_1} \ . \label{xiIasp}
\end{eqnarray}
We see that slow roll is not guaranteed close to the fixed point $\phi_0$ and
that we have to assume further that $\beta_1 \ll 1$. This case is also special
from the holographic point of view to be discussed in the next section.

\begin{figure}[h]
\begin{center}
\epsfig{file=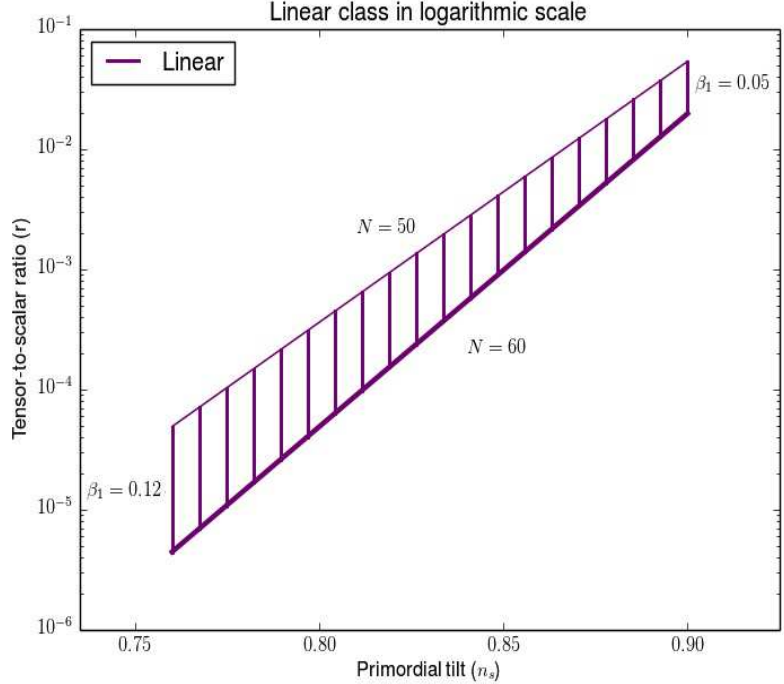,width=12cm,height=10cm}
\end{center}
\caption{Models of the linear class ({\bf Ia(1)}) in the plot ($(r,n_s$):
values of $\beta_1$ (see (\ref{betaIasp})) range from $0.05$ (top right) to
$0.12$ (bottom left). Each
vertical segment corresponds to values of $N$ ranging from $50$ (top) to
$60$ (bottom). The maximal relative error is $8 \cdot10^{-3}$.}
\label{figlinear}
\end{figure}

We show in Figure \ref{figlinear} the typical region covered by this class of
models in the plot $r$ vs $n_s$. The observable parameters $n_s,r$ depend on
two model parameters, $\beta_1,N$. On the other hand, $\alpha_s=(n_s-1)r/ 8$
is not independent.

\subsection{Large field inflation ({\bf Ib})} \label{Ib}

The second class ({\bf Ib}) is similar to the previous one
except that {\em the fixed point is
reached at $\phi \rightarrow + \infty$}. It thus falls into what is usually
called the large field inflation. The leading term of the $\beta$-function is now
\begin{equation}
\label{betaIb}
\beta(\phi) \simeq - {\hat \beta_p \over \left[\kappa\phi \right]^p}
\ .
\end{equation}

\vskip .3cm
$\bullet$ {\bf Inverse monomial class: Ib(p)}, $p > 1$
\vskip .3cm
We consider first the case $p \in R$, $p > 1$:
\begin{eqnarray}
W(\phi) &=& W_0 \exp \left\{- {\hat \beta_p \over 2 (p-1)}{1 \over
\left[\kappa\phi \right]^{p-1}}
\right\} \ , \label{WIb} \\
V(\phi) &=& {3 W_0^2 \over \kappa^2} \left[1 -  {\hat \beta_p \over (p-1)}
{1 \over \left[\kappa\phi \right]^{p-1}} + {\cal O} \left(\phi^{-2(p-1)}\right)
\right] \ , \label{VIb}
\end{eqnarray}
We then obtain for the number of e-foldings
\begin{equation}
N =  {\left[\kappa\phi \right]^{p+1}\over (p+1) \hat \beta_p} - \lambda \ ,
\ \ \ \  \lambda \equiv {\left[\kappa\phi_f \right]^{p+1}\over (p+1) \hat
\beta_p} \ . \label{NIb}
\end{equation}
Assuming as usual $\left| \beta(\phi_f) \right| \sim 1$, we have $\lambda
\sim \hat \beta_p^{1/p}/(p+1)$. The $\beta$-function can be expressed in terms of
$N$:
\begin{equation}
\label{betaNIb}
\beta(N) = - {\hat \beta_p^{\frac{1}{p+1}} \over \left[(p+1) (N+\lambda)
\right]^{\frac{p}{p+1}}} \ .
\end{equation}

We have for the scalar spectral index:
\begin{eqnarray}
\label{nSIb}
n_s -1 &\sim&  - 2 {\beta' \over \kappa} = -2p{\hat \beta_p \over
 \left[\kappa\phi \right]^{(p+1)}}\sim - {2p\over p+1} {1 \over N} \ , \\
\label{aSIb}
\alpha_s  &\sim& -2 {\beta \beta''\over \kappa^2} = -2p(p+1) {\hat
\beta_p^2 \over  \left[\kappa\phi \right]^{2(p+1)}} \sim -{2p \over p+1}{1
\over N^2} \ ,
\end{eqnarray}
whereas the tensor to scalar ratio reads
\begin{equation}
\label{rIb}
r = 8 \beta^2 \sim {8 \hat \beta_p^{2/(p+1)}
\over \left[ (p+1)  N\right]^{2p/(p+1)}} \ .
\end{equation}
We note that this class of models corresponds to $\alpha = 2p/(p+1)$ in the
notations of (\ref{Mukh1}). Since $p>1$, $1<\alpha<2$.

For completeness, the slowroll parameters of the inverse
monomial class are:
\begin{equation}
\label{reltoNIb}
\eta_H = - {p\over p+1} {1 \over N} \ , \ \ \ \epsilon_H = {\hat
\beta_p^{2/(p+1)} \over 2 \left[(p+1) N\right]^{2p/(p+1)}} \ , \ \ \
\xi_H^2 = {p \over p+1}{1 \over N^2} \ .
\end{equation}

\begin{figure}[h]
\begin{center}
\epsfig{file=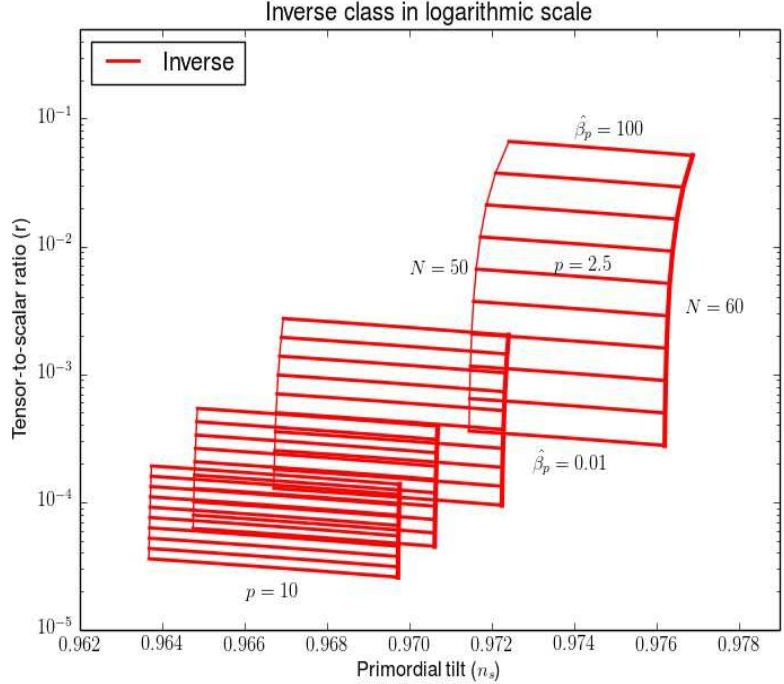,width=12cm,height=10cm}
\end{center}
\caption{Models of the inverse class ({\bf Ib(p)}, $p>1$) in the plot ($(r,
n_s$): values of $p$ are from right to left to $=2.5,5,7.5,10$, and
values of $\hat \beta_p$ (see (\ref{betaIb})) range from $10^{-2}$ (bottom) to
$10^2$ (top) in logarithmic spacing. Each horizontal segment corresponds to
values of $N$ ranging from $50$ (left) to $60$ (right). The maximal relative
error is $9 \cdot10^{-3}$.}
\label{figinverse}
\end{figure}

We show in Figure \ref{figinverse} the typical region covered by this class of
models in the plot $r$ vs $n_s$. As in the {\bf Ia(q)} class,  the three
experimentally accessible parameters, $n_s,  \alpha_s, r$, depend on three
other parameters, namely $q,N $ and $\beta_q$. One may thus think that one
can adjust independently the three cosmological parameters, but this is not
necessarily compatible with our choice of perturbativity i.e. restricting to
the first term (\ref{betaIb}) in the expansion of the beta function.

\vskip .3cm
$\bullet$ {\bf Chaotic class: Ib(1)},
\vskip .3cm
A special case is obtained for $p=1$. The superpotential reads
\begin{equation}
\label{WIbsp}
W(\phi) = W_0 \left[\kappa\phi \right]^{\hat \beta_1/2} \ ,
\end{equation}
where $C$ is a constant. One obtains the potential of chaotic inflation
\cite{Linde:1983gd}:
\begin{equation}
\label{VIbsp}
V(\phi) = {3 W_0^2 \over 4 \kappa^2} \left( 1 - {\hat \beta_1^2 \over
6 \left[\kappa\phi  \right]^2}\right) \left[\kappa\phi
\right]^{\hat \beta_1} \sim {3 W_0^2 \over 4 \kappa^2} \left[\kappa\phi
\right]^{\hat \beta_1} \ ,
\end{equation}
Note that, in this special case, the power is given by the  $\beta$-function
coefficient. The number of e-foldings is:
\begin{equation}
N =  {\left[\kappa\phi \right]^2 \over 2 \hat \beta_1}- \lambda \ ,
\ \ \ \lambda \equiv {\left[\kappa\phi_f \right]^2 \over 2 \hat \beta_1}\ ,
\label{NIbsp}
\end{equation}
where $\lambda \sim \beta_1 /2$, if $\left|\beta(\phi_f)\right| \sim 1$. Thus
\begin{equation}
\label{betaNIbsp}
\beta(N) \sim - \left({\beta_1 \over 2(N+ \lambda)}\right)^{1/2} \ ,
\end{equation}
which shows that this model corresponds to $\alpha = 1$.
We thus have for the scalar spectral index and the tensor to scalar ratio:
\begin{eqnarray}
\label{nSIbsp}
n_s -1 &\sim& -{\hat \beta_1 \over \left[\kappa\phi  \right]^2} (2+\hat \beta_1)
\sim -{1 + \hat \beta_1/2 \over N} \ , \\
\label{aSIbsp}
\alpha_s &\sim&  -{\hat \beta_1^2 \over \left[\kappa\phi  \right]^4}
(2\hat \beta_1 + 4) \sim -(1 + \hat \beta_1/2){1 \over N^2} \ , \\
\label{rIbsp}
r &=& 8 \beta^2 \sim {4 \hat \beta_1 \over N} \ .
\end{eqnarray}

For reference, the slowroll parameters of the chaotic
class read:
\begin{equation}
\label{reltoNIbsp}
\epsilon_H = {\hat \beta_1 \over 4N} \ ,\ \ \
\eta_H = {\hat \beta_1 -2 \over 4N} \ ,  \ \
\xi_H^2 =  {(\hat \beta_1 -2)(\hat \beta_1 -4) \over 16 N^2} \ ,
\end{equation}

\begin{figure}[h]
\begin{center}
\epsfig{file=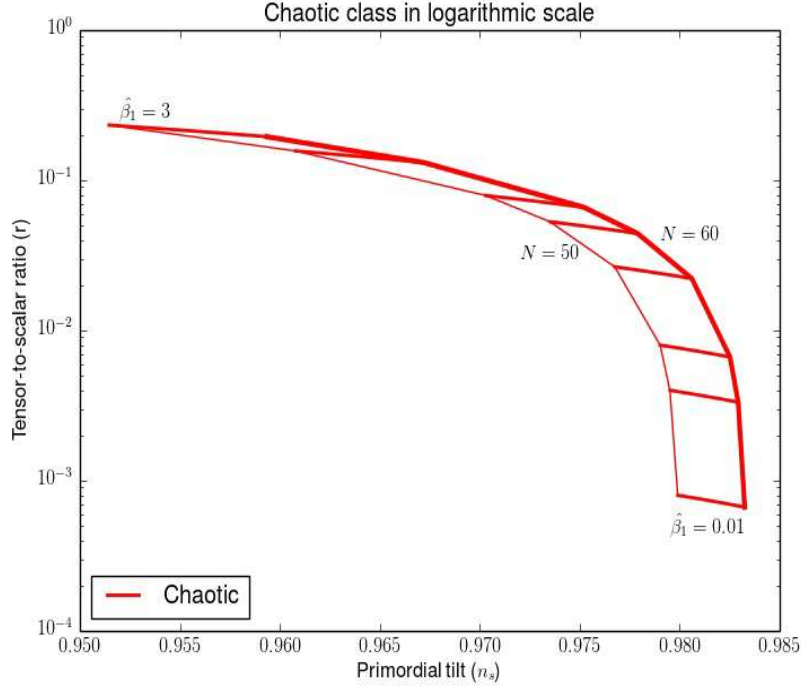,width=12cm,height=10cm}
\end{center}
\caption{Models of the chaotic class ({\bf Ib(1)}, $p=1$) in the plot ($(r,
n_s$): values of $\hat \beta_1$ (see (\ref{betaIb})) are from bottom to
top $=0.01,0.05,0.1,1/3,2/3,1,2,3$.
Each horizontal segment corresponds to
values of $N$ ranging from $50$ (left) to $60$ (right). The maximal relative
error is $8 \cdot10^{-4}$}
\label{figchaotic}
\end{figure}

We show in Figure \ref{figchaotic} the typical region covered by this class of
models in the plot $r$ vs $n_s$. In this subclass, the observable parameters
$n_s,r$ depend on two parameters of the model, $\beta_1$ and $N$. On the other
hand, $\alpha_s = -(1- n_s)(1-n_s-r/8)$ is not independent.

\begin{figure}[h]
\begin{center}
\epsfig{file=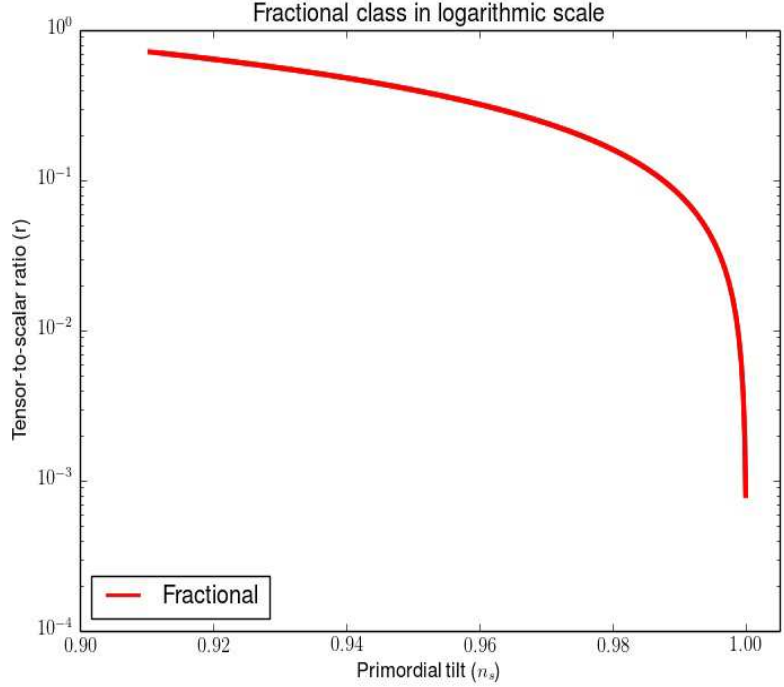,width=12cm,height=10cm}
\end{center}
\caption{Models of the fractional class ({\bf Ib(p)}, $0<p<1$) in the plot ($(r,
n_s$): values of $p$ range from $0.001$ (right) to $0.005$ (left)  and
values of $\hat \beta_p$ (see (\ref{betaIb})) range from $0.001$ (bottom right)
to $0.3$ (top left). This corresponds to a maximal relative of $5 \cdot10^{-3}$. }
\label{figfractional}
\end{figure}

\vskip .6cm
$\bullet$ {\bf Fractional class: Ib(p)}, $0 < p < 1$
\vskip .5cm
Finally, for $0<p<1$, the superpotential reads:
\begin{equation}
\label{WIb2}
W = W_0 \exp \left[ {\hat \beta_p \over 2(1-p)}\left[\kappa\phi
\right]^{1-p}\right] \ ,
\end{equation}
where $W_0$ is a constant, with corresponding potential
\begin{equation}
\label{VIb2}
V \sim {3W_0^2 \over 4\kappa^2} \exp \left[ {\hat \beta_p \over 1-p}
\left[\kappa\phi \right]^{1-p}\right] \ .
\end{equation}
Then (\ref{NIb}) and (\ref{betaNIb}) remain valid but, this time, one
has\footnote{Correspondingly, the slowroll parameters of the power law class
satisfy:
\begin{equation}
\label{reltoNIb2}
\epsilon_H = \eta_H = \xi_H = \frac{1}{2} \beta(\phi)^2
= {\hat \beta_p^{2/(p+1)} \over 2 [(p+1)N]^{{2p\over p+1}}} \ .
\end{equation}
Their leading contribution cancels in (\ref{dnSdlnkH}) and one thus has to go
to subleading order to obtain (\ref{aSIb2}) from the slowroll parameters.}
\begin{eqnarray}
\label{nSrIb2}
1- n_s = {r \over 8} &=& \beta^2 = {\hat \beta_p^{2/(p+1)} \over
[(p+1)N]^{{2p\over p+1}}} \ , \\
\label{aSIb2}
\alpha_s  &=& -2{\beta^2 \beta' \over \kappa} =
- 2p {\hat \beta_p^{{2\over (p+1)}} \over
[(p+1)N]^{{3p+1\over p+1}}} \ .
\end{eqnarray}
This case, represented in the plot $r$ vs $n_s$ on Figure \ref{figfractional},
corresponds to $\alpha = 2p/(p+1)$ with $0<\alpha<1$. In this class, the tensor
to scalar ratio and scalar index are related through $1- n_s = {r \over 8}$
(as can be seen from the Figure), while $\alpha_s$ can be adjusted
independently.

\vskip .3cm
$\bullet$ {\bf Power law class: Ib(0)}
\vskip .3cm

The limit $p \rightarrow 0$ of the previous class is of a special type because
it lies outside the strict framework of our analysis: the beta function tends
to a constant i.e. $\beta(\phi) = -\hat \beta_0$ for $\phi \rightarrow \infty$,
which corresponds to power law inflation \cite{Lucchin:1984yf}, as we will now
see.

In this case we have indeed
\begin{eqnarray}
\label{WIb0}
W(\phi) &=& W_0 e^{{\hat \beta_0\over 2}\kappa\phi} \ , \\
\label{VIb0}
V(\phi) &=& {W_0^2 \over 8 \kappa^2} \left(6-\hat \beta_0^2\right)
e^{\hat \beta_0\kappa\phi} \ .
\end{eqnarray}
In this case, one can easily check that the scale factor evolves as
\begin{equation}
\label{a(t)Ib0}
a(t) \sim t^{2/\hat \beta_0^2} \ ,
\end{equation}
which corresponds to power law inflation. As is well-known, this model is
incomplete since there is no end to inflation: another regime has to take over
at some later time.

We have
\begin{equation}
\label{NIb0}
N = {1 \over \hat \beta_0} \left( \kappa \phi - \kappa \phi_0\right) \ ,
\end{equation}
and\footnote{The slowroll parameters are just obtained from (\ref{reltoNIb2})
by taking the limit $p \rightarrow 0$: $\epsilon_H = \eta_H = \xi_H =
\frac{1}{2} \hat \beta_0^2$.}
\begin{equation}
\label{miscIb0}
n_s - 1 = -{\hat \beta_0^2 \over 1-\hat \beta_0^2/2} \ \ \ , \ \
r = 8 \hat \beta_0^2 \ \ \ , \ \ {dn_s \over d \ln k} = 0 \ ,
\end{equation}
where the latter relation signals exact scaling.
Assuming $\hat\beta_0 \ll 1$, we have, as in the previous case
\begin{equation}
\label{relIb0}
r=-8(n_s-1)
\end{equation}

This class, has a single parameter $\hat \beta_0$, that controls all critical
exponents and is therefore the most  ``predictive" of all classes. It is
interesting to note that it accommodates both the parameters constrained by
Planck as well as a value of $r$ in the $10^{-1}$ range.

\subsection{Exponential class ({\bf II})} \label{II}

In a third class of models ({\bf II($\gamma$)}), the fixed point is again
reached for
$\phi \rightarrow \infty$ (large field), but, as in the case $\alpha = 2$ of
section \ref{sect:1}, the expansion is in terms of
$Y \equiv \exp (-\gamma \kappa \phi)$, $\gamma$ being a positive constant,
\cite{Kiritsis}.  If the first term is of order $Y^n$, we may redefine $\gamma$
in order to include $n$ into it. We may thus write the  $\beta$-function
(\ref{beta}) as
\begin{equation}
\label{betaII}
\beta (\phi) = -\beta Y \ ,
\end{equation}
which corresponds to
\begin{eqnarray}
W(\phi) &=& W_0 \exp \left\{- {\beta \over 2 \gamma}  Y  \right\} \ ,
\label{WII} \\
V(\phi) &=& {3 W_0^2 \over 4 \kappa^2} \left[1 - {1 \over \gamma} \beta Y
+ {\cal O}\left( Y^{2} \right) \right] \ . \label{VII}
\end{eqnarray}
Starobinsky model (\cite{Starobinsky:1980te}) and Higgs inflation
\cite{Bezrukov:2007ep} belong to this class,
which corresponds to $\alpha = 2$ in Mukhanov classification (compare with
(\ref{Mukh20W}) and (\ref{Mukh20})). We then obtain the number of e-foldings:
\begin{equation}
\label{NII}
N =  {1 \over  \gamma \beta Y}-{1 \over  \gamma \beta Y_f}
\sim -{1 \over \gamma \beta(\phi)} - {1 \over \gamma}  \ ,
\end{equation}
where we have assumed $\left|\beta(\phi_f)\right| \sim 1$. The  $\beta$-function
then reads in terms of $N$:
\begin{equation}
\label{betaNII}
\beta(N) \sim -{1 \over \gamma N+1} \ .
\end{equation}
We thus have
for the scalar spectral index and the tensor to scalar ratio:
\begin{eqnarray}
\label{nSII}
n_s -1 &\sim&- 2 {\beta' \over \kappa}= -{2 \over N} \ , \\
\label{aSII}
\alpha_s &\sim& - 2 {\beta \beta'' \over \kappa^2} = -{2 \over N^2}
\ , \\
\label{rII}
r &\sim& 8 \beta^2 = {8 \over  \gamma^2 N^2} \ .
\end{eqnarray}
For reference, the slowroll parameters of the
exponential class read:
\begin{equation}
\label{reltoNII}
\epsilon_H = {1 \over 2 \gamma^2 N^2} \ , \ \
\eta_H = - {1\over N} \ , \ \
\xi_H = { 1 \over N} \ .
\end{equation}

\begin{figure}[h]
\begin{center}
\epsfig{file=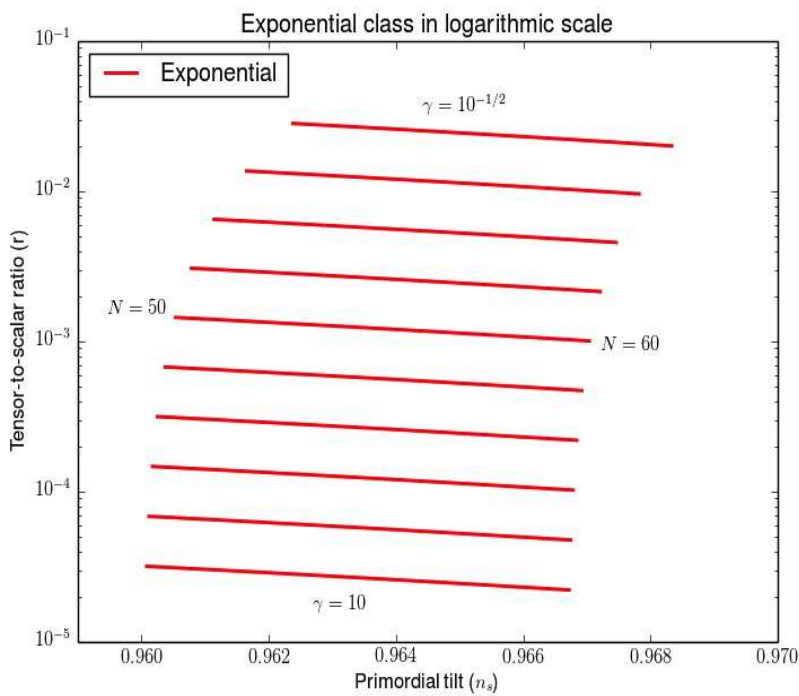,width=12cm,height=10cm}
\end{center}
\caption{Models of the exponential class ({\bf II($\gamma$)}) in the plot ($(r,
n_s$): values of $\gamma$ range from
$10^{-1}$ (bottom) to $10$ (top)  in logarithmic spacing and
each horizontal segment corresponds to values of $N$ ranging from $50$ (left)
to $60$ (right). The maximal error is $4 \cdot10^{-3}$.}
\label{figexp}
\end{figure}

We show in Figure \ref{figexp} the typical region covered by this class of
models in the plot $r$ vs $n_s$. The observable parameters ($n_s,r,\alpha_s$)
depend only on two numbers, $\gamma, N$. In particular,
$\alpha_s=-(n_s-1)^2/2$.

\begin{figure}[t]
\begin{center}
\epsfig{file=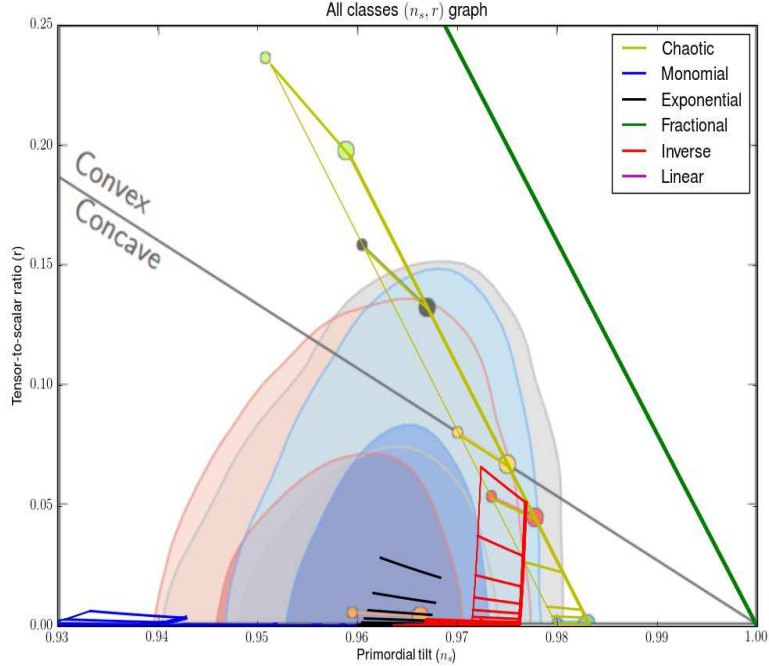,width=12cm,height=10cm}
\end{center}
\caption{Single field models of inflation in the plane $(n_s,r)$:
class {\bf Ia} with $q=1$ (linear), $q>1$
(monomial); class {\bf Ib} with $0<p<1$ (fractional), $p=1$
(chaotic), $p>1$ (inverse); class {\bf II} (exponential). Values favoured
by Planck data \cite{Ade:2013uln} are also indicated.}
\label{fig1}
\end{figure}

\begin{figure}[t]
\begin{center}
\epsfig{file=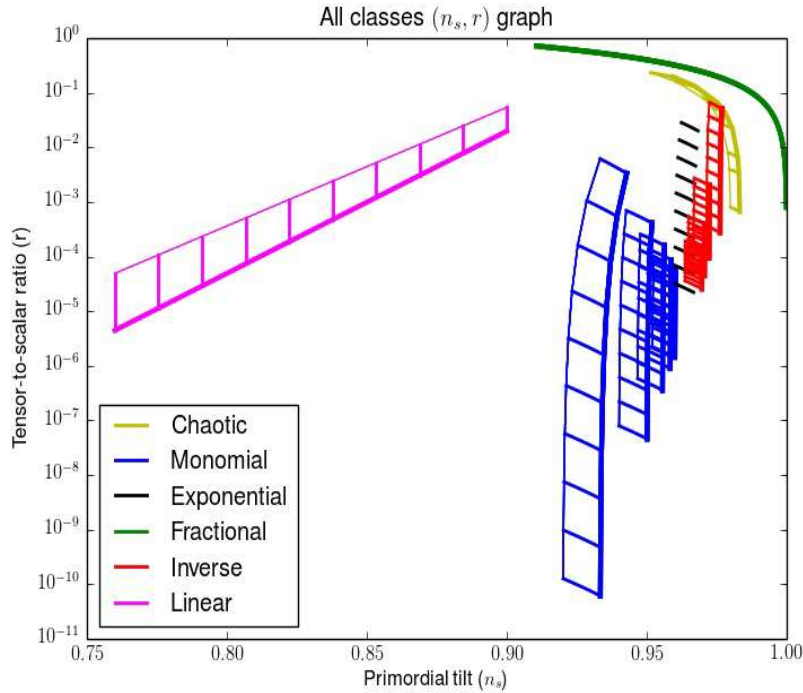,width=12cm,height=10cm}
\end{center}
\caption{Same as Figure \ref{fig1} but in logarithmic scale for $r$ which shows
the behaviour for small values of $r$.}
\label{fig2}
\end{figure}

\vskip .5cm
To summarize, Figures \ref{fig1} and \ref{fig2} present the various classes of
models in a
plot $(n_s,r)$. Figure \ref{fig1} shows how each class of models fares with
respect to the constraints imposed by Planck data. Each model
is represented by a segment corresponding to values of $N$ from $50$ (left end)
to $60$ (right end). This is just indicative and the segment can be extended to the
right for larger values of $N$. Figure \ref{fig2} shows in
more details the region of low values for $r$ and how the different classes
converge for low values of $r$.

\begin{figure}[t]
\begin{center}
\epsfig{file=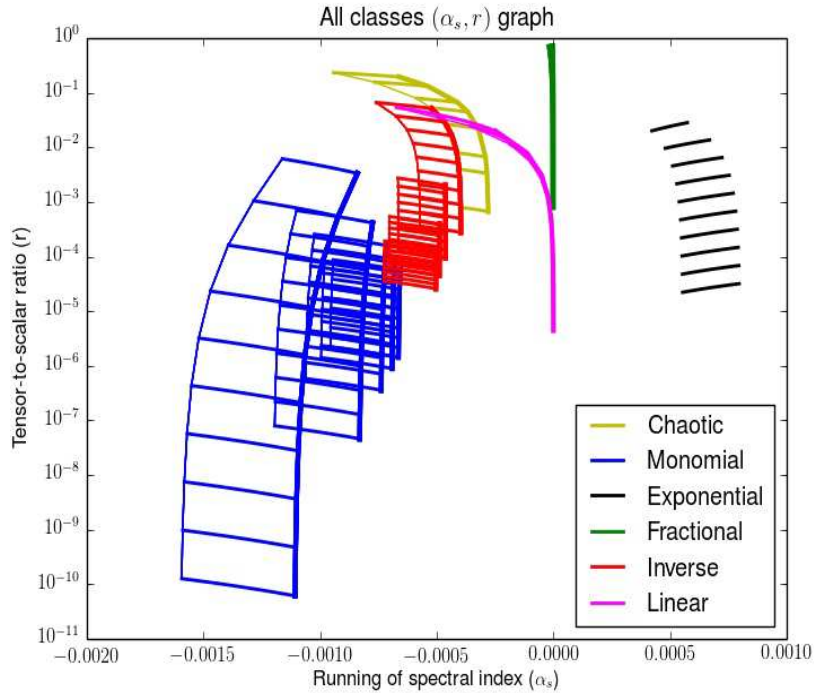,width=12cm,height=10cm}
\end{center}
\caption{Variation of $r$ versus the running parameter $\alpha_s = dn_S/d\ln k$
for the various classes of models}
\label{fig3}
\end{figure}

We also show in Figure \ref{fig3} how the parameter $r$ varies with respect to
the running of the scalar index $\alpha_s$.

\section{Renormalisation group and the AdS/CFT correspondence}
\label{sect:4}
\setcounter{equation}{0}

The appearance of a renormalisation group equation in this cosmological context
is not fortuitous \cite{McFadden,Kiritsis}. Indeed,
inflationary scenarios can be viewed as departures from a pure de Sitter
solution and it has been noted by many authors that de Sitter space can be
mapped to an Anti de Sitter space which may be equivalently described by a
conformal field theory through the famous AdS/CFT correspondence
\cite{Maldacena}.  Departure from de Sitter is therefore translated into
departure from conformal invariance, which is described as usual by a
renormalisation group equation. One may thus use this correspondence to
identify the field theories equivalent to the inflation scenarios. Issues like
the naturalness of the initial conditions or of the choice of potential may
receive a new light in this context. Moreover, because the AdS/CFT
correspondence relates strongly (resp. weakly) coupled gravities with weakly
(resp. strongly) coupled field theories, this may also lead to new theories of
inflation  \cite{McFadden}.

More explicitly, the correspondence between de Sitter (dS) and anti-de Sitter
(AdS) gravity theory (see \cite{McFadden,Kiritsis} for details)
may be performed by taking $m_P^2$ into $-m_P^2$ and $V(\phi)$ into $-V(\phi)$.
The Hubble constant $H$ on the dS side corresponds to $L^{-1}$, where
$L$ is the AdS curvature length scale, and the time
coordinate in de Sitter becomes the AdS holographic spacelike coordinate $u$.
Using the standard Poincar\'e coordinates\footnote{where the metric takes the
form: $ds^2 = {L^2 \over u^2}  \left[du^2 + dx_a dx^a \right]$.} for AdS,
$u \rightarrow \infty$ corresponds to the horizon, whereas $u \rightarrow 0$
corresponds to the boundary of AdS spacetime. This boundary corresponds in dS
(in the flat slicing generally used to desribe inflation) to the cosmic scale factor
becoming infinitely large.

AdS/CFT correspondence states that a gravity theory on AdS is equivalent to a
Conformal Field Theory (CFT) on its boundary (hence with one dimension less,
i.e. three in the case we consider; one then writes AdS$_4$/CFT$_3$). The holographic
coordinate $u$ may be seen as
corresponding to the energy scale of the boundary field theory: the horizon
$u \rightarrow + \infty$ corresponds to the low energy (infrared) limit
whereas the high energy (ultraviolet) limit corresponds to $u \rightarrow 0$,
i.e. towards the boundary itself.
Thus, as we move along the holographic direction parametrized by $u$, we go
from a small cosmic scale factor (on the dS gravity side) corresponding to the
infrared limit of the
CFT ($u \rightarrow \infty$) to a exponentially large cosmic scale factor
corresponding to the ultraviolet regime of the CFT ($u \rightarrow 0$).
Hence the de Sitter fixed point in the neighborhood of which we start to inflate
corresponds to an infrared fixed point in the dual quantum field theory.

As we have seen in Section \ref{sect:1}, inflation corresponds to a slight
departure from de Sitter geometry (represented by a zero of the  $\beta$-function
(\ref{beta})). In the holographic description, this means that we must depart
from a strictly conformal field theory. Such a departure is described by an
operator ${\cal O}(x)$ characterized by a scaling dimension $\Delta$: on the
holographic dual
field theory, the nontrivial renormalisation group flow is induced by the
scaling behaviour of this operator (if $\Delta > d$, where $d=3$ is the dimension
of the boundary, the perturbation induced by ${\cal O}(x)$ is called irrelevant; if
$\Delta < d$, it is relevant and, if $\Delta = d$, it is marginal). This operator
is thus dual to the inflaton
field, and classification of such operators leads to a classification of
inflation theories. Indeed, inflation corresponds to flow in the scaling regime in the neighborhood of the fixed point.

Following the standard AdS$_4$/CFT$_3$ transcription, the field
$\phi(u,x)$ in the bulk of AdS$_4$ (from which one can obtain the field in dS$_4$) is
related to the
source field $\phi_0(x)$ on the boundary, conjugate to the field theory operator
${\cal O}(x)$, by the following boundary condition:
\begin{equation}
\label{boundary}
\phi(u,x) = u^{3-\Delta} \phi_0(x) + u^\Delta \langle {\cal O} \rangle + \cdots, \
\ \ \ u \rightarrow 0 \ .
\end{equation}
We finally note that the scaling dimension $\Delta$ is directly computed to the
mass $m$ of the scalar field through the relation:
\begin{equation}
\label{dimension-mass}
\Delta  = {3 \over 2} + {1 \over 2} \sqrt{9 + 4m^2 L^2} \ .
\end{equation}
In the cosmological case, the AdS curvature scale $L$ is replaced by the
inverse of the de Sitter Hubble parameter, $H_i^{-1}$.

An important ingredient in the correspondence is the expected relation between
power spectra in cosmology and correlators in QFT \cite{smf}.
Starting from the cosmological expressions of the scalar and tensor power
spectra defined as
\be
\Delta_{s}^2(p)\equiv {p^3\over 2\pi^2}\langle \zeta(p)\zeta(-p)\rangle ={p^3
\over 2\pi^2}|\zeta_p(0)|^2
\label{ds}\ee
\be
\Delta_{t}^2(p)\equiv {p^3\over 2\pi^2}\langle \gamma_{ij}(p)
\gamma_{ij}(-p)\rangle ={2p^3\over \pi^2}|\gamma_p(0)|^2
\label{gs}\ee
where $\zeta_p(0)$ and $\gamma_p(0)$ are the constant late-time values of the
cosmological mode functions of the curvature, $\zeta_p(t)$, and the
transverse-traceless part of the metric, $\gamma_p(t)$.
On the other hand in QFT$_3$ we have the correlator of the stress tensor
\be
\langle T_{ij}(p)T_{kl}(-p)\rangle =T(p)\Pi_{ijkl}+S(p)\pi_{ij}\pi_{kl}
\ee
where $\Pi$ is the 3d transverse-traceless projector
\be
\Pi_{ijkl}\equiv {1\over 2}(\pi_{ik}\pi_{jl}+\pi_{il}\pi_{jk}-\pi_{ij}\pi_{kl})\sp
\pi_{ij}=\delta_{ij}-{p_ip_j\over p^2}
\ee
The correspondence maps the power spectra to the analytically continued QFT
spectral densities as \cite{smf}
\be
\Delta_{s}^2(p)=-{p^3\over 16\pi^2~Im~S(-ip)}\sp \Delta_{t}^2(p)=-{2p^3\over
\pi^2~Im~T(-ip)}
\ee

A scalar QFT operator with scaling dimension $\Delta$ therefore gives
\be
\Delta_{s}^2(p)\sim p^{2(3-\Delta)}
\ee
Matching this with the definition (in cosmology) $\Delta_s^2\sim p^{n_s-1}$ we
obtain
\be
\Delta =3+{1\over 2}(1-n_s)
\ee

\subsection{Class ({\bf Ia})}

From what has been said above, this class naturally divides into two subclasses,
depending on whether the squared mass $m^2$ of the scalar field at the fixed
point is vanishing, in which case one has a marginal operator ($\Delta = 3$),
or negative\footnote{As explained already in Section \ref{Ia}, the case of
positive mass squared ($0<q<1$) leads to infinite mass at the fixed point, and
thus no inflation.}, in which case ones has an irrelevant operator
($\Delta > 3$).
\vskip .3cm
$\bullet$ Marginal operators ({\bf  Ia(q)}, $q > 1$)
\vskip .3cm
The case with $q>1$ and integer maps to marginal interactions, where the first
non-trivial contributions to the $\beta$-function appear at $q-1$ loop order.
Such a  case  has been analyzed in detail in \cite{Bourdier} in the
context of holography.\footnote{No known such case is known to our knowledge in string theory.}

The case where q is larger than one  but real can be thought of as a case where the leading
contribution to the  $\beta$-function is non-analytic in the coupling. Although
this can happen in CFT (and it happens in thermal gauge theories), we are not
aware of a case where the leading contribution is non-analytic but this cannot
be excluded.

\vskip .3cm
$\bullet$ Slightly irrelevant operator ({\bf  Ia(1)})
\vskip .3cm
In this case, the squared mass of the scalar field is obtained, on the dS gravity side,
from (\ref{VIasp}):
\begin{equation}
\label{mIasp}
m^2 H_i^{-2} = -\beta_1 (3+\beta_1) \ ,
\end{equation}
where we have used $W_0 = - 2H_i$. Comparing with what is obtained on the dual field
theory from (\ref{dimension-mass}):
\begin{equation}
\label{mIaspdual}
m^2 L^2 = \Delta (\Delta -3) \ ,
\end{equation}
(and accounting for a sign when going from dS to AdS) we conclude that
\begin{equation}
\label{beta1}
\beta_1 = \Delta -3 \ .
\end{equation}
The condition $\beta_1 \ll 1$ imposes to choose $\Delta \simeq 3$ which corresponds to
a slightly irrelevant operator, that is an operator of dimension $\Delta \gsim 3$.

\subsection{Class ({\bf Ib})}

 This is described by the $\beta$ function in (\ref{betaIb}).
It can be thought of as a strong-weak coupling dual of the previous case {\bf Ia}.
Indeed the data computed through it for the most part are analytic continuations of {\bf Ia}. Such cases are known in QFT,  the most famous being N=1 SQCD in the conformal window where the dual magnetic theory has weak coupling when the parent theory has strong coupling. There,  the coupling and the dual coupling roughly flow to IR fixed points so that one is the strong coupling dual of the other

For example defining a new "coupling", $\kappa\chi={1\over \kappa \phi}$ the problem now is equivalent to a new scalar sector
\begin{equation}
S_{scalar}\sim \int d^4x\sqrt{g}\left[{1\over 2}G(\chi)(\partial\chi)^2-\tilde V(\chi)\right]
\end{equation}
where the metric\footnote{This is known as the Zamolodchikov metric in QFT.}  $G(\chi)$ is non-trivial, and singular at the deSitter point.
In this example with $\chi={1\over \phi}$
\begin{equation}
G(\chi)={1\over \chi^4}\sp \tilde V(\chi)=V\left(1\over \chi\right)\simeq  3 {H_i^2
\over \kappa^2} \left[1 -  {\hat \beta_p \left[\kappa\chi \right]^{p-1} \over (p-1)
} + \cdots \right] \ , \label{VIb}
\end{equation}
with  the de Sitter solution at $\chi=0$ where the Zamolodchikov metric $G(\chi)$ is singular.

A subclass of {\bf Ib} is interesting in its own right. It is {\bf  Ib(1)}.
This corresponds to $V\sim e^{a\phi}$ with $a\phi\to +\infty$. This class is not associated with a CFT (deSitter solution) but with a scaling theory that violates hyperscaling, \cite{Charmousis}.

This leads to power-law inflation.

Note that if a potential is diverging faster than $e^{\sqrt{3\over 2}\phi}$ as $\phi\to \infty$ then this region is not penetrable by the equations of motion. This implies that a divergence like $e^{\phi^p}$ with $p>1$ is impossible.

In string theory such exponential behavior is ubiquitous as the Zamolodchikov metric of many scalars is logarithmic
\begin{equation}
S_{scalar}\sim \int d^4x\sqrt{g}\left[{1\over 2}{(\partial T)^2 \over T^2}-C T^b\right]
\end{equation}
which upon redefinition gives exponential potentials.

\subsection{Class II}

This is the class with asymptotically flat potentials named AFIM in \cite{Kiritsis}.
Its holographic map corresponds to a $\beta$ function of the form,
\be
\beta\simeq b_0 Y^2+{\cal O}(Y^3)\sp Y\to 0
\ee
Interpreting $Y$ as a coupling constant\footnote{Indeed in a string theory context that gauge coupling constant is the exponential of the dilaton scalar.}, this corresponds to the asymptotically free class of QFTs like QCD, \cite{Kiritsis}. Such a class of potentials was used to model holographic models for YM theory in four dimensions, \cite{ihqcd}.
This class has a single parameter, $\gamma$.

\section{Analysing more general models in terms of the characteristic classes}
\label{sect:5}
\setcounter{equation}{0}

The identification of fundamental classes allows to analyze more complex
inflation models in terms of these building blocks. Indeed, one can build
out of the classes presented in the last sections, models that
interpolate between two different classes. This is usually done by introducing
a new scale $f$.
When this scale is small, i.e. $(\kappa f)^2 \ll 1$, one is in
a small field regime, when the scale is large,
i.e. $(\kappa f)^2 \gg 1$, one is in a large field regime.
For the cases discussed in this paper the large field limit typically falls in class {\bf Ia} and the small field limit corresponds to the chaotic
class. Interestingly enough, when the scale is intermediate, one may still
obtain the slow roll conditions for inflation.

The way this works in general is that any $\beta$-function can be made small
enough for inflation to occur if multiplied by a small-enough parameter.
For example consider an arbitrary function $f(\phi)$ that is bounded.
The $\beta$-function $\beta(\phi)=f(\phi)$ would describe inflation near the
zeros of $f(\phi)$. However, a $\beta$-function, $\beta(\phi)=x f(\phi)$ with
$x\ll 1$ will inflate for all $\phi$ so that $x f(\phi)$ is small. In this way
for sufficiently small $x$, any $\beta$-function can lead to sufficient
inflation. As it will be seen below, all such cases can be mapped via fields
redefinitions to the following action
\begin{equation}
\label{action}
{\cal S}= -\int d^4x \sqrt{-g} \left[ m_{_P}^2 \left({1\over 2} R
+ V(\tilde\phi)\right) +{ f^2 \over 2}
\partial^\mu \tilde \phi \partial_\mu \tilde\phi \right] \ .
\end{equation}
where now $\tilde\phi$ is dimensionless and $f\gg m_P$.
Such a gravitational theory leads to inflation for generic bounded potentials
$V(\tilde\phi)$.

However, experience from gravity and string theory indicates that this
``fine-tuning" of the gravitational action is probably unrealizable in a
healthy gravitational theory. The argument is similar to that in \cite{AH}.
The scalar mediates a much weaker force than gravity although its charges are
similar to those of gravity. It is expected that such a context will lead to
inconsistencies with black-holes both in asymptotically flat but also in an
asymptotically AdS context. This argument needs to be made precise but we
will not pursue this further here. We will merely note that there is no known
such case in string theory.

\begin{figure}[t]
\begin{center}
\epsfig{file=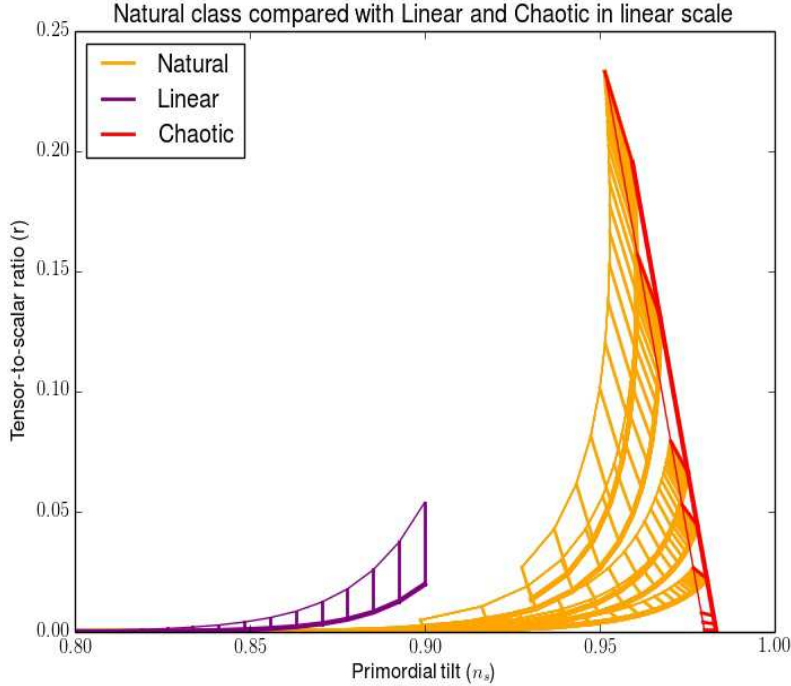,width=12cm,height=10cm}
\end{center}
\caption{Models of generalized natural inflation (in yellow) interpolating
between the
linear case (in purple) and chaotic inflation (in red),
in the plot ($n_s$,$r$). }
\label{fig4}
\end{figure}

We illustrate this with models inspired by natural inflation
\cite{Binetruy:1986ss,Freese:1990rb} which turn out to interpolate between the
linear class and the chaotic class (see Figure \ref{fig4}), as we now show.
Our starting point is the beta function
\begin{equation}
\label{Abeta}
\beta(\phi) = (\kappa f)^{-1} \tan (\phi / 2f) \ .
\end{equation}

The corresponding superpotential is $W(\phi) = W_0 \cos (\phi/2f)$, with $W_0$
constant, and the potential reads
\begin{equation}
\label{Apot}
V = {3W_0^2 \over 8 \kappa^2}\left(1-{1 \over 6(\kappa f)^2}\right)
+ {3W_0^2 \over 8 \kappa^2}\left(1+{1 \over 6(\kappa f)^2}\right) \cos{\phi
\over f} \ .
\end{equation}
This coincides with the potential of natural inflation:
\begin{equation}
\label{Apotnatural}
V = {V_0 \over 2} \left( 1 + \cos {\phi \over f} \right) \ ,
\end{equation}
with $V_0 \equiv 3W_0^2/(4\kappa^2)$, in the limit\footnote{Note that the
minimum at $\phi/f = \pi$ does not appear as a
fixed point of the  $\beta$-function (\ref{Abeta}). Accordingly, the two potentials
differ at this point, but this is negligible in the limit (\ref{Acondition}).}
\begin{equation}
\label{Acondition}
(\kappa f)^2 \gg 1/6 \ .
\end{equation}
We have
\begin{equation}
\label{AN}
N = -(\kappa f)^2 \ln {\sin^2\left( {\phi \over 2f}\right) \over
\sin^2\left( {\phi_f \over 2f}\right)} \ ,
\end{equation}
where the value of $\phi_f$ of the field at the end of inflation is given by
the condition $\beta(\phi_f) =1$:
\begin{equation}
\label{Aphif}
\sin^2(\phi_f/2f)=\frac{(\kappa f)^2}{1+(\kappa f)^2} \ .
\end{equation}
We can express the  $\beta$-function in terms of the number $N$ of e-foldings:
\begin{equation}
\label{Abeta(N)}
\beta(N(\phi))=\frac{1}{\sqrt{(1+(\kappa f)^2)
\exp\left\{\frac{N}{(\kappa f)^2}\right\}-(\kappa f)^2}}\ .
\end{equation}
This expression is useful to discuss the two regimes of inflation.

If $(\kappa f)^2 \ll 1$, then we can write the  $\beta$-function as
\begin{equation}
\label{Abeta(N)1}
\beta(N)=\exp\left\{-\frac{N}{2(\kappa f)^2}\right\} \ ,
\end{equation}
where we recognize a  $\beta$-function of the linear class {\bf Ia} with $q=1$
and $\beta_1 = \frac{1}{2(\kappa f)^2}$ (see (\ref{betaNIasp}). Indeed this
corresponds to the small field regime where $\phi/2f \ll 1$ and (\ref{Abeta})
reads
\begin{equation}
\label{Abeta1}
\beta(\phi) \sim {(\kappa \phi) \over 2 (\kappa f)^2} \ .
\end{equation}

The other regime corresponds to $N \ll (\kappa f)^2$ in which case
\begin{equation}
\label{Abeta(N)2}
\beta(N) = {1 \over \sqrt{N+1}} \ ,
\end{equation}
where we recognize the  $\beta$-function of the chaotic class {\bf Ib} with $p=1$
and $\beta_1 = 2$.

In order to see this from the form of the  $\beta$-function in terms of the
scalar field, one has to redefine $\phi$ and consider instead $\phi' \equiv
f\pi -\phi$, in the limit $\phi'/f \rightarrow 0$ or $\phi/f \rightarrow \pi$.
Then
\begin{equation}
\label{Abeta2}
\beta(\phi')=\frac{1}{\kappa f}\tan\left(\frac{\pi}{2}-\frac{\phi'}{2f}\right)
=\frac{1}{\kappa f}\cot\left(\frac{\phi'}{2f}\right)
=\frac{1}{\kappa f}\frac{1}{\phi'/2f}=\frac{2}{\kappa\phi'}.
\end{equation}
End of inflation happens for $\phi_f$ close to $f \pi$ as can be seen from
(\ref{Aphif}), or $\phi_f' \ll f$. The fixed point is still at $\phi = 0$, that
is $\phi' = f\pi$.

We thus see that natural inflation interpolates between the linear class  ({\bf
Ia}, q=1) and the chaotic class ({\bf Ib}, p=1). More generally, it is possible
to interpolate between a linear model and any chaotic model by choosing the
 $\beta$-function in the following way:
\begin{equation}
\label{Abetagen}
\beta (\phi) = (\kappa f)^{-1} \tan \left(\phi/\lambda f \right) \ ,
\end{equation}
in which case one obtains a chaotic model with parameter $\beta_1 = \lambda$.
  This is seen explicitly in
Figure \ref{fig4}, which shows how generalized natural inflation models merge
into our (linear and chaotic) classes of models.

One may further understand how the two regimes (linear and chaotic) appear
in natural inflation. In order to do this, let us write the potential
(\ref{Apot}) in terms of $N$ using (\ref{AN}) and (\ref{Aphif}):
\begin{equation}
\label{AVN}
V\left(N(\phi)\right) = V_0 \left( 1-\left({(\kappa f)^2 +1/6 \over 1 + (\kappa
f)^2}\right)\exp\left[ -{N(\phi) \over \kappa f^2} \right] \right) \ .
\end{equation}

\begin{figure}[t]
\begin{center}
\epsfig{file=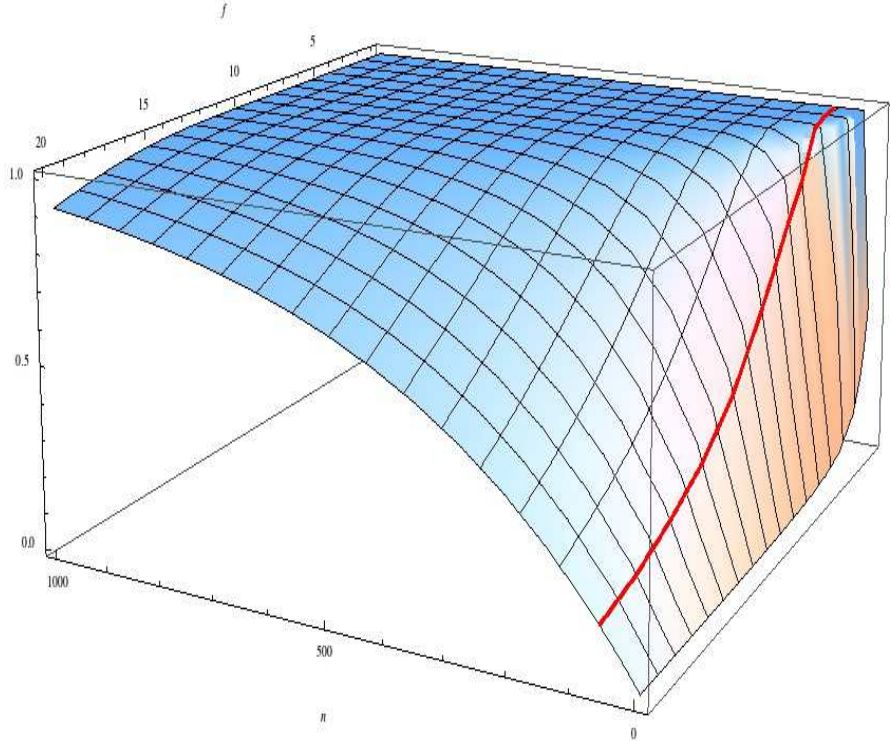,width=12cm,height=10cm}
\end{center}
\caption{Potential $V\left(N(\phi)\right)$ of the generalized natural
inflation type
(\ref{Apot}) as a function of $N$, $1<N<1000$, for values of
$f$ in the range $1< \kappa f <20$.}
\label{figpot}
\end{figure}

In Figure \ref{figpot}, we represent this potential for various values of
$\kappa  f$. Remember that, with our definition (\ref{Nbeta}) of $N$, $N$
vanishes
at the end of inflation whereas $N$ increases when one reaches more and more
primordial times. In Figure \ref{figpot}, the red curve represents $N=60$. We see that the low $\kappa f$ regime corresponds to
the plateau region associated with the region of the potential associated with
region close to the maximum (as in the linear class of inflation models). But
as soon as one gets to higher values of $\kappa f$, one leaves the vicinity of
the maximum and reaches the intermediate region of the potential, where chaotic
inflation takes over.

\section{Conclusions} \label{sect:6}
\setcounter{equation}{0}

At a time where the physics of inflation becomes data oriented, it is important
to withdraw from specific model analyses and to attempt to interpret the data
in terms of large classes. This is what we have proposed in this work, based on
(A)dS/CFT correspondence. We have shown that this does make sense from the
point
of view of the gravity theory alone, in the context of the Hamilton-Jacobi
approach. What the AdS/CFT correspondence brings in addition is the fact that
the classes that we define are {\em universality} classes in the sense of
quantum  field theories or condensed matter, with associated critical indices.
 Also, the perspective given by the field theory being different from the one
adopted on the gravity theory side, the different classes may be natural
in different ways. Moreover, the AdS/CFT correspondance being of a strong/weak
duality, one may address the question of finding new strongly coupled
gravity theories for inflation \cite{McFadden}.

Such a map may have far-reaching consequences both for the issues of
singularities in cosmology as well as the fate of the many massive scalars
that are present in most fundamental gravitational theories.
There is however an important puzzle that is associated with expanding
universes. A contracting universe  is evolving in accordance with QFT
renormalization: from the UV at early times to the IR at late times.
In contrast, for an expanding universe the cosmological evolution is opposite
in direction to the RG flow in a QFT: it runs from the IR at early times to
the UV at late times.
This puzzle is sharpened in the context of mirage cosmology \cite{mirage}, where
also its potential resolution can be glimpsed.
 In that setup, an expanding cosmology corresponds to a brane-universe moving away from a gravitation stack of $D_3$ branes, therefore against the gravitational attraction. For this to happen the universe brane must have some energy, and in \cite{mirage,review} it was advocated that this could happen for a brane universe in an elliptical path around a central stack of D$_3$ branes or in an unbounded outbound path.
 In the first case it would lead to a cyclic cosmological evolution on the universe brane, alternatively approaching and moving away from the central stack. A concrete model of cosmological evolution  along these lines was developed in \cite{gk}. In the second case it leads to a standard cosmology where the universe expands for ever.

Among the generalisations that we have not considered
in this paper is the generalization to multi-field inflation
\cite{Salopek:1990jq,Bourdier}, to fields with nontrivial kinetic terms
(and $c_s\not = 1$), as well as the computation of non-gaussianities in these
different classes.

\section{ Acknowledgements}

We would like to thank F. Bouchet, M. M\"{u}nchmeyer, F. Nitti and
S. Sarkar for useful discussions.

We acknowledge the financial support of the UnivEarthS
Labex program at Sorbonne Paris Cit\'e (ANR-10-LABX-0023 and
ANR-11-IDEX-0005-02).
This work was also supported in part by European Union's Seventh Framework Programme under grant agreements (FP7-REGPOT-2012-2013-1) no 316165,
PIF-GA-2011-300984, the EU program ``Thales'' MIS 375734, by the European Commission under the ERC Advanced Grant BSMOXFORD 228169 and was also co-financed by the European Union (European Social Fund, ESF) and Greek national funds through the Operational Program ``Education and Lifelong Learning'' of the National Strategic Reference Framework (NSRF) under ``Funding of proposals that have received a positive evaluation in the 3rd and 4th Call of ERC Grant Schemes''.

 \newpage
\appendix

 \renewcommand{\theequation}{\thesection.\arabic{equation}}
\addcontentsline{toc}{section}{Appendices}
\section*{APPENDIX}

\section{General formulas} \label{App:A}
\setcounter{equation}{0}

We assemble in this appendix complete formulae which express the cosmological parameters in terms of the beta function and its derivatives.
\begin{itemize}
	\item Number of e-foldings:
	  	\begin{align}	
       			 N(\phi)=-\kappa\int_{\phi_{f}}^\phi\frac{\mathrm{d}\phi'}{\beta(\phi')}.
 		\end{align}
	\item Superpotential:
		\begin{align}	
    			 W=W_{0}\exp\left\{-\frac{\kappa}{2}\int\mathrm{d}\phi\beta(\phi)\right\}.
 		\end{align}	
	\item Potential:
		\begin{align}	
    			 V&=\frac{3}{4\kappa^2}W^2\left(1-\frac{\beta^2}{6}\right),\\
    			 V_{,\phi}&=\frac{3}{4\kappa^2}W^2\left(-\kappa\beta\left(1-\frac{\beta^2}{6}\right)-\frac{\beta\beta_{,\phi}}{3}\right),\\
  			 V_{,\phi\phi}&=\frac{3}{4\kappa^2}W^2\left(\kappa^2\beta^2\left(1-\frac{\beta^2}{6}\right)-\kappa\beta_{,\phi}\left(1-\frac{5\beta^2}{6}\right)-\frac{1}{3}\left(\beta_{,\phi}^2+\beta\beta_{,\phi\phi}\right)\right),\\
  			 V_{,\phi\phi\phi}&=\frac{3}{4\kappa^2}W^2\left(-\kappa^3\beta^3\left(1-\frac{\beta^2}{6}\right)+3\kappa^3\beta\beta_{,\phi}\left(1-\frac{\beta^2}{2}\right) \right. \notag\\
			& \left. +\frac{\kappa}{3}\beta\left(6\beta_{,\phi}^2+\beta\beta_{,\phi\phi}\right)-k\beta_{,\phi\phi}\left(1-\frac{5\beta^2}{6}\right)-\beta_{,\phi}\beta_{,\phi\phi}-\frac{1}{3}\beta\beta_{,\phi\phi\phi}\right).
		\end{align}
	\item Slow-roll parameters:
	  	\begin{align}
       			 \epsilon_H&\equiv\frac{2}{\kappa^2}\left(\frac{H'}{H}\right)^{2}=\frac{1}{2}\beta^2,\\
			 \eta_H&\equiv\frac{2}{\kappa^2}\frac{H''}{H}=\frac{1}{2}\beta^2-\frac{\beta_{,\phi}}{\kappa},\\
			 \xi_H^2&\equiv\frac{4}{\kappa^4}\left(\frac{H'H'''}{H^2}\right)=\frac{1}{4}\beta^4-\frac{3}{2\kappa}\beta^2\beta_{,\phi}+\frac{1}{\kappa^2}\beta\beta_{,\phi\phi}.
 		\end{align}
	\item Scalar power spectrum
		\begin{align}
			\label{beta-scalar-spectra}
			 \mathcal{P}_s(k)&=\kappa^2\frac{k^3}{2\pi^2}\left|\frac{v_k}{a\beta}\right|^2,
		\end{align}
		where $u_k$ is computed from:
		\begin{align}
		\label{eqm-scalar-spectra}
			 &\frac{v_{k,\phi\phi}}{\kappa^2}+\frac{1}{\beta}\left(1-\frac{\beta^2}{2}+\frac{\beta_{,\phi}}{\kappa}\right)\frac{v_{k,\phi}}{\kappa}+\notag\\&+\frac{4}{\beta^2}\left(\frac{k^2}{a^2W^2}-\frac{1}{2}\left[1-\frac{\beta^2}{4}+\frac{3}{2}\frac{\beta_{,\phi}}{\kappa}-\frac{\beta^2}{4}\frac{\beta_{,\phi}}{\kappa}+\frac{1}{2}\left(\frac{\beta_{,\phi}}{\kappa}\right)^2+\frac{\beta^2}{2}\frac{\beta_{,\phi\phi}}{\kappa^2}\right]\right)v_k=0.
		\end{align}
	\item Tensor power spectrum
		\begin{align}
		\label{beta-tensor-spectra}
			 \mathcal{P}_t(k)&=8\kappa^2\frac{k^3}{2\pi^2}\left|\frac{u_k}{a}\right|^2,
		\end{align}
		where $v_k$ is computed from:
		\begin{align}
		\label{eqm-tensor-spectra}
			 &\frac{u_{k,\phi\phi}}{\kappa^2}+\frac{1}{\beta}\left(1-\frac{\beta^2}{2}+\frac{\beta_{,\phi}}{\kappa}\right)\frac{u_{k,\phi}}{\kappa}+\frac{4}{\beta^2}\left(\frac{k^2}{a^2W^2}-\frac{1}{2}+\frac{\beta^2}{8}\right)u_k=0.
		\end{align}
		
\end{itemize}
Eq.\eqref{eqm-scalar-spectra} and eq.\eqref{eqm-tensor-spectra} allow us to
compute the exact power spectra. Their solutions can be found with numerical
methods. Such a computation however lies beyond the scope of this paper and
will be addressed in future work.

To plot the predictions for the observable quantities (see Figures \ref{figmonomial} to \ref{figexp}) we have proceeded as follows:
\begin{itemize}
\item Given the standard expressions for the scalar and tensor power spectra:
\begin{equation}
\mathcal{P}_{s}(k)=\frac{1}{4\pi^2}\left.\frac{H^4}{\dot{\phi}^2}\right|_{k=aH} \qquad \qquad \mathcal{P}_{t}(k)=8\left.\left(\frac{H}{2\pi}\right)^2\right|_{k=aH}
\end{equation}
we express them using typical quantities of our formalism:
 \begin{equation}
\mathcal{P}_{s}(k)=\frac{1}{4\pi^2}\left.\frac{W_{,\phi}^2}{\beta^4}\right|_{k=\frac{aW_{,\phi}}{\beta}} \qquad \qquad \mathcal{P}_{t}(k)=\frac{2}{\pi^2}\left.\left(\frac{W_{,\phi}}{\beta} \right)^2\right|_{k=\frac{aW_{,\phi}}{\beta}}
\end{equation}
\item We compute the observables quantities of interest for us i.e.
 \begin{equation}
n_{s}= -\frac{\left( \beta^2 + 2 \beta_{,\phi} \right)}{\left( 1 -\beta^2 \right)}   \qquad \qquad n_{t} = -\frac{ \beta^2 }{\left( 1 -\beta^2 \right)}  \qquad \qquad \dots
\end{equation}
\item We identify the leading terms for all of these expressions.
\item We neglect all of those terms expected to produce higher order
contributions. In this procedure we keep terms whose numerical value is at
least 1\% of the leading one. For example $\lambda$ (e.g. defined in
Eq.\eqref{equ:lambdadef})
is of order one and cannot be neglected (as we did in the main text) when
compared to $N\sim 50-60$ (see e.g. Eq.(\ref{betaNIa})): to produce the
figures contained in this paper, we kept it.
\item We compare the numerical results shown in the figures with the ones including next order correction. The highest value of the relative error for each class is reported in the corresponding figure caption.
\end{itemize}

\section{Hilltop inflation: interpolating between the monomial and the
chaotic class}
\label{App:B}
\setcounter{equation}{0}

\begin{figure}[h]
\begin{center}
\epsfig{file=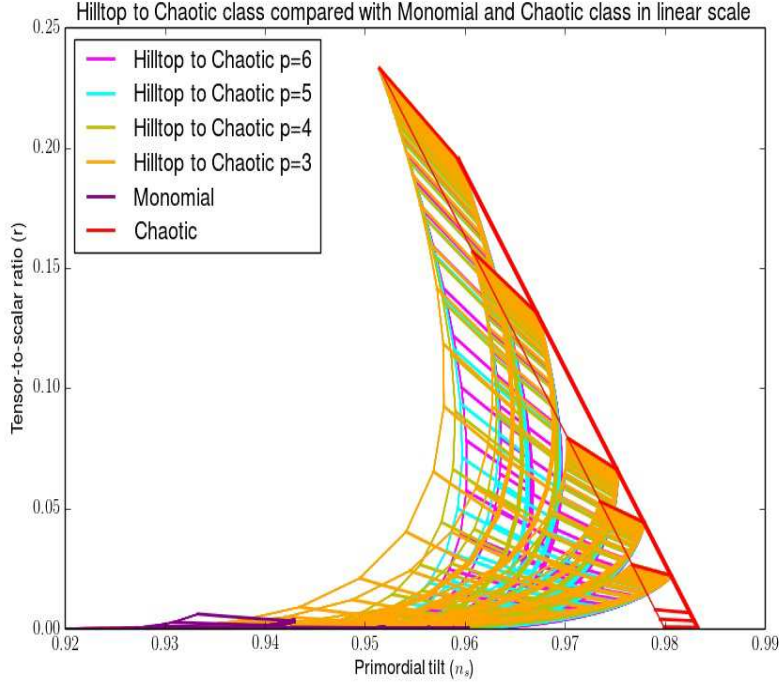,width=12cm,height=10cm}
\end{center}
\caption{Generalized hilltop models interpolating between the monomial class
and the chaotic class.}
\label{fig5}
\end{figure}

We consider in this Appendix the  case of the so-called generalized
hilltop potentials, which, as we know, interpolates between the monomial
and the chaotic classes (see Figure \ref{fig5}).
Let us consider the beta function:
\begin{equation}
\label{hilltopbeta}
\beta (\phi) = {\beta_p \over \kappa f} {p(\phi/f)^{p-1} \over 1-(\phi/f)^p} \ .
\end{equation}
with $p>2$. Inflation takes place in the range $0<\phi<f$, the fixed point
being at $\phi=0$.

The superpotential is $W(\phi)=W_0(1-(\phi/f)^p)^\frac{\beta_p}{2}$, with
$W_0$ constant, and the potential reads:
	\begin{equation}
	\label{pothtc}	
			V=V_0\left(1-\frac{1}{6}\left(\frac{\beta_p p}{\kappa f}\right)^2\frac{(\phi/f)^{2(p-1)}}{(1-(\phi/f)^p)^2}\right)\left(1-(\phi/f)^p\right)^{\beta_p}.
	\end{equation}
where $V_0=\frac{3W_0^2}{4}$
The number of e-foldings is given by:
		\begin{equation}	
		\label{Nefhtc}
			N(\phi)=\frac{(\kappa f)^2}{p\beta_p}\left[\frac{1}{(p-2)}\left((\phi/f)^{2-p}-(\phi_f/f)^{2-p}\right)+\frac{1}{2}\left((\phi/f)^{2}-(\phi_f/f)^{2}\right)\right].
		\end{equation}
	where $\phi_f$ is the value of the field at the end of inflation given by the condition $\beta(\phi_f)=1$.  However, this relation cannot be solved analytically for a general value of $p$. Moreover, writing $\phi$ as a function of $N$ requires to invert equation \eqref{Nefhtc}, which is also not possible analytically. To find the predictions of the general model, numerical methods are used. Nevertheless, we investigate the limiting cases, for respectively the small and large field. \\

\paragraph{Small field limit}
	In the limit $(\phi/f)\ll1$, we can write the $\beta$-function as:
		\begin{equation}
		\label{betahtcsf}
			\beta(\phi)=\frac{p\beta_p}{\kappa f}(\phi/f)^{p-1}=\frac{p\beta_p}{(\kappa f)^p}(\kappa\phi)^{p-1}.
		\end{equation}
	We recognize the $\beta$-function of the monomial class {\bf Ia}  with $q\equiv p-1$ and $\beta_q\equiv\frac{p\beta_p}{(\kappa f)^p}$. When $p>2\ (q>1)$, it corresponds to hilltop models. In this limit the potential is:
		\begin{equation}	
		\label{pothtcsf}
			V\simeq V_0\left(1-\beta_p(\phi/f)^p\right),
		\end{equation}
 and the number of e-foldings:
		\begin{equation}	
		\label{Nefhtcsf}
			N(\phi)\simeq\frac{(\kappa f)^2}{p\beta_p}\frac{1}{(p-2)}\left((\phi/f)^{2-p}-(\phi_f/f)^{2-p}\right).
		\end{equation}
	We define the constant $\lambda\equiv\frac{(\kappa f)^2}{p\beta_p}\frac{1}{(p-2)}(\phi_f/f)^{2-p}$, the condition $\beta(\phi_f)=1$ gives:
		\begin{equation}
		\label{lambdahtcsf}
			\lambda=\frac{\kappa \phi_f}{p-2}=\frac{1}{p-2}\left(\frac{(\kappa f)^p}{p\beta_p}\right)^{\frac{1}{p-1}}.
		\end{equation}
We can write $\beta(\phi)$ in terms of $N$:
		\begin{equation}
		\label{betaNhtcsf}	
       				\beta(N)\simeq\frac{1}{\left[\left(\frac{p\beta_p}{(\kappa f)^p}\right)^\frac{1}{p-1}(p-2)(N+\lambda)\right]^\frac{p-1}{p-2}}.
		\end{equation}

\paragraph{Large field limit}
	In the limit $(\phi/f)\rightarrow1$, we redefine $\phi$ and consider $\phi'\equiv f-\phi$, in the regime $(\phi'/f)\rightarrow0$. The $\beta$-function becomes:
		\begin{equation}
		\label{betahtclf}	
			\beta(\phi')=\frac{\beta_p}{\kappa f}\frac{p(1-(\phi'/f))^{p-1}}{1-(1-(\phi'/f))^p}\simeq\frac{\beta_p}{\kappa f}\frac{p(1-(p-1)(\phi'/f))}{1-(1-p(\phi'/f))}\simeq\frac{\beta_p}{\kappa f}\frac{1}{(\phi'/f)}=\frac{\beta_p}{\kappa\phi'}.
		\end{equation}
	This is the $\beta$-function corresponding to the chaotic class {\bf 1b}, with $p=1$. It corresponds to a large field model, which is achieved with $f\gg1$. Note that the fixed point is still at $\phi=0$ corresponding to $\phi'=f$.
	The potential can be approximated as:
		\begin{equation}
		\label{pothtclf}		
			V\simeq V_0\left(p(\phi'/f)\right)^{\beta_p}.
		\end{equation}
	The number of e-foldings in this limit becomes:
		\begin{equation}
		\label{Nefhtclf}	
			 N(\phi)\simeq\frac{1}{2\beta_p}\left[(\kappa\phi')^2-(\kappa\phi_f')^2\right].
		\end{equation}
	Assuming the end of inflation for $\beta(\phi'_f)=1$ we get:
		\begin{equation}
		\label{endhtclf}	
       			\phi'_f/f=\beta_p/(\kappa f),
		\end{equation}
	and we express the $\beta$-function in terms of $N$:
		\begin{equation}
		\label{betaNhtclf}
       			\beta(N)\simeq\sqrt{\frac{\beta_p}{2N+\beta_p}}.
		\end{equation}

\par
In conclusion, using a single $\beta$-function, we interpolate between any
hilltop model to any chaotic one. The parameter $p$ in the $\beta$-function
determines the hilltop limit, whereas the parameter $\beta_p$ fixes the
chaotic side.



\addcontentsline{toc}{section}{References}
\begin{thebibliography}{99}

\def\hri#1#2{\href{http://arxiv.org/abs/#1}{[ArXiv:#1]#2}}
\def\hre#1#2{\href{http://arxiv.org/abs/#1/#2}{[ArXiv:#1/#2]}}
\def\hspi#1#2{\href{http://www.slac.stanford.edu/spires/find/hep/www?irn=#1}{#2}}

\bibitem{Ade:2013uln}
  P.~A.~R.~Ade {\it et al.}  [Planck Collaboration],
  {\em ``Planck 2013 results. XXII. Constraints on inflation,''}
  \hri{1303.5082}{[astro-ph.CO]}.

  \bibitem{BICEP}
  P.~A.~R.~Ade {\it et al.}  [BICEP2 Collaboration],
  {\em ``BICEP2 I: Detection Of B-mode Polarization at Degree Angular Scales,''}
  \hri{1403.3985}{[astro-ph.CO]}.

\bibitem{Starobinsky:1980te}
  A.~A.~Starobinsky,
  Phys.\ Lett.\ B {\bf 91} (1980) 99.

\bibitem{Bezrukov:2007ep}
  F.~L.~Bezrukov and M.~Shaposhnikov,
  Phys.\ Lett.\ B {\bf 659} (2008) 703
  [arXiv:0710.3755 [hep-th]].

\bibitem{Mukhanov:2013tua}
  V.~Mukhanov,
  Eur.\ Phys.\ J.\ C {\bf 73} (2013) 2486
  [arXiv:1303.3925 [astro-ph.CO]].

    \bibitem{Roest}
  D.~Roest,
  {\em ``Universality classes of inflation,''}
  JCAP {\bf 01} (2014) 007
  \hri{1309.1285}{[hep-th]}.

  \bibitem{GB}
  J.~Garcia-Bellido and D.~Roest,
  {\em ``The large-N running of the spectral index of inflation,''}
  \hri{1402.2059}{[astro-ph.CO]}.

\bibitem{Kiritsis}
  E.~Kiritsis,
  {\em ``Asymptotic freedom, asymptotic flatness  and cosmology,''}
  JCAP {\bf 1311} (2013) 011
  \hri{1307.5873}{[hep-th]}.


    \bibitem{mirage}
  A.~Kehagias and E.~Kiritsis,
  {\em ``Mirage cosmology,''}
  JHEP {\bf 9911} (1999) 022
  \hre{hep-th}{9910174}.

     \bibitem{strominger}
  A.~Strominger,
  {``The dS / CFT correspondence,''}
  JHEP {\bf 0110} (2001) 034
  \hre{hep-th}{0106113}.

\bibitem{McFadden}
  P.~McFadden and K.~Skenderis,
  {\em ``Holography for Cosmology,''}
  Phys.\ Rev.\ D {\bf 81} (2010) 021301
  \hri{0907.5542}{[hep-th]}.

   \bibitem{review}
  E.~Kiritsis,
  {\em ``D-branes in standard model building, gravity and cosmology,''}
  Phys.\ Rept.\  {\bf 421} (2005) 105
   [Erratum-ibid.\  {\bf 429} (2006) 121]
   [Fortsch.\ Phys.\  {\bf 52} (2004) 200]
  \hre{hep-th}{0310001}.

  \bibitem{gk}
  C.~Germani, N.~E.~Grandi and A.~Kehagias,
  {\em ``A Stringy Alternative to Inflation: The Cosmological Slingshot Scenario,''}
  Class.\ Quant.\ Grav.\  {\bf 25} (2008) 135004
  \hre{hep-th}{0611246};\\
  {\em ``The Cosmological Slingshot Scenario: Myths and Facts,''}
  Gen.\ Rel.\ Grav.\  {\bf 42} (2010) 77
  \hri{0706.0023}{[hep-th]};\\
C.~Germani, N.~Grandi and A.~Kehagias,
  {\em ``The Cosmological Slingshot Scenario: A Stringy Proposal for the Early Time Cosmology,''}
  AIP Conf.\ Proc.\  {\bf 1031} (2008) 172
  \hri{0805.2073}{[hep-th]}.







\bibitem{Salopek:1990jq}
  D.~S.~Salopek and J.~R.~Bond,
  Phys.\ Rev.\ D {\bf 42} (1990) 3936.

\bibitem{Lidsey:1995np}
  J.~E.~Lidsey, A.~R.~Liddle, E.~W.~Kolb, E.~J.~Copeland, T.~Barreiro and
M.~Abney,
  Rev.\ Mod.\ Phys.\  {\bf 69} (1997) 373
  [astro-ph/9508078].

\bibitem{Binetruy:2006ad}
  P.~Bin\'etruy,
  ``Supersymmetry: Theory, experiment and cosmology,''
  Oxford, UK: Oxford Univ. Pr. (2006) 520 p

\bibitem{Binetruy:1986ss}
  P.~Bin\'etruy and M.~K.~Gaillard,
  Phys.\ Rev.\ D {\bf 34} (1986) 3069.

\bibitem{Freese:1990rb}
  K.~Freese, J.~A.~Frieman and A.~V.~Olinto,
  Phys.\ Rev.\ Lett.\  {\bf 65} (1990) 3233.



\bibitem{Linde:1983gd}
  A.~D.~Linde,
  Phys.\ Lett.\ B {\bf 129} (1983) 177.





\bibitem{Maldacena}
  J.~M.~Maldacena,
  Adv.\ Theor.\ Math.\ Phys.\  {\bf 2} (1998) 231
  [hep-th/9711200].

\bibitem{Stewart:1993bc}
  E.~D.~Stewart and D.~H.~Lyth,
  Phys.\ Lett.\ B {\bf 302} (1993) 171
  [gr-qc/9302019].


\bibitem{Bourdier}
  J.~Bourdier and E.~Kiritsis,
  {\em ``Holographic RG flows and nearly-marginal operators,''}
  Class.\ Quant.\ Grav.\  {\bf 31} (2014) 035011
  \hri{1310.0858}{[hep-th]]}.


\bibitem{Charmousis}
  C.~Charmousis, B.~Gouteraux, B.~S.~Kim, E.~Kiritsis and R.~Meyer,
  {\em ``Effective Holographic Theories for low-temperature condensed matter systems,''}
  JHEP {\bf 1011} (2010) 151
  \hri{1005.4690}{[hep-th]};\\
  B.~Gouteraux and E.~Kiritsis,
  {\em ``Generalized Holographic Quantum Criticality at Finite Density,''}
  JHEP {\bf 1112} (2011) 036
  \hri{1107.2116}{[hep-th]}.



  \bibitem{ihqcd}
  U.~Gursoy and E.~Kiritsis,
  {\em ``Exploring improved holographic theories for QCD: Part I,''}
  JHEP {\bf 0802} (2008) 032
\hri{0707.1324}{[hep-th]};\\
  U.~Gursoy, E.~Kiritsis and F.~Nitti,
  {\em ``Exploring improved holographic theories for QCD: Part II,''}
  JHEP {\bf 0802} (2008) 019
  \hri{0707.1349}{[hep-th]}.

  \bibitem{T}
  U.~Gursoy, E.~Kiritsis, L.~Mazzanti and F.~Nitti,
  {\em ``Holography and Thermodynamics of 5D Dilaton-gravity,''}
  JHEP {\bf 0905} (2009) 033
  \hri{0812.0792}{[hep-th]}.

  \bibitem{smf}
  P.~McFadden and K.~Skenderis,
  {\em ``The Holographic Universe,''}
  J.\ Phys.\ Conf.\ Ser.\  {\bf 222} (2010) 012007
  \hri{1001.2007}{[hep-th]}.

\bibitem{AH}
  N.~Arkani-Hamed, L.~Motl, A.~Nicolis and C.~Vafa,
  {\em ``The String landscape, black holes and gravity as the weakest force,''}
  JHEP {\bf 0706} (2007) 060
  \hre{hep-th}{0601001}.






\end{thebibliography}
\end{document}